\documentclass[conference]{IEEEtran}
\usepackage{url}
\usepackage[pdftex]{graphicx}
\pdfoutput=1 

\usepackage[margin=1.5cm]{geometry}
\usepackage{array, xcolor}
\usepackage{lipsum}
\usepackage{bibentry}
\usepackage{longtable}
\usepackage{authblk}
\usepackage{times}
\usepackage{helvet}
\usepackage{courier}
\usepackage{color}
\usepackage{soul}
\usepackage{url}
\usepackage{xcolor}
\usepackage{stfloats}
\usepackage{caption}
\usepackage{tabularx}
\usepackage{epstopdf}
\usepackage{rotating}
\usepackage{epstopdf}
\usepackage{siunitx}
\usepackage{booktabs}
\usepackage{subcaption}
\usepackage{tabularx}
\usepackage{subcaption}
\usepackage{array}
\usepackage{amsmath}

\renewcommand{\vec}[1]{\mathbf{#1}}

\begin{document}


\title{Tracking Urban Activity Growth Globally with Big Location Data}


\author[1]{Matthew L Daggitt}
\author[2]{Anastasios Noulas}
\author[3]{Blake Shaw}
\author[1]{Cecilia Mascolo}

\affil[1]{Computer Laboratory, University of Cambridge, UK}
\affil[2]{Data Science Institute, Lancaster University, UK}
\affil[3]{Twitter Inc., USA}

\maketitle

\begin{abstract}
In recent decades the world has experienced rates of urban growth unparalleled in any other period of history and this growth is shaping the environment in which an increasing proportion of us live. 
In this paper we use a longitudinal dataset from Foursquare, a location-based social network, to analyse urban growth across 100 major cities worldwide. 

Initially we explore how urban growth differs in cities across the world. We show that there exists a strong spatial correlation, with nearby pairs of cities more likely to share similar growth profiles than remote pairs of cities.
Subsequently we investigate how growth varies inside cities and demonstrate that, given the existing local density of places, higher-than-expected growth is highly localised while lower-than-expected growth is more diffuse.
Finally we attempt to use the dataset to characterise  competition between new and existing venues.
By defining a measure based on the change in throughput of a venue before and after the opening of a new nearby venue, we demonstrate which venue types have a positive effect on venues of the same type and which have a negative effect.
For example, our analysis confirms the hypothesis that there is large degree of competition between bookstores, in the sense that existing bookstores normally experience a notable drop in footfall after a new bookstore opens nearby.
Other place categories however, such as Airport Gates or Museums, have a cooperative effect and their presence fosters higher traffic volumes to nearby places of the same type.
\end{abstract}

\section{introduction}
\label{sec:intro}
Urban growth has been a focal point for global development in recent decades. Large cities are often characterised as the steam engines of the world's economy and they have given rise to a new inter-disciplinary field: \textit{urban data science}.
This new science of cities~\cite{batty2013new} is powered by the increasing availability of new layers of data describing human activity in urban settings. 
The analysis and modelling solutions typically proposed in this context have focused on human mobility and transport modelling~\cite{noulas2012tale}, neighbourhood detection~\cite{cranshaw2012livehoods}\cite{ zhang2013hoodsquare} and urban scaling~\cite{bettencourt2013origins}\cite {arcaute2013city} amongst others. 

However, advances in city science have not only come from work primarily concerned with classical urban geography and ecology problems, but have also been driven by computer science and related mobile technologies that lie at the centre of today's digital revolution.
Mapping and sensing technologies, alongside mobile crowdsourcing applications, services and systems have enabled a powerful feedback loop between data-driven solutions and problems that arise in the urban domain. 
Location-based services in particular have played a vital role, with Foursquare being one of the best examples of this new paradigm. 
Foursquare is powered by more than $60$ million users who crowdsource information about places, neighbourhoods and cities as they move. More than $80$ thousand applications rely on, and contribute to, its data ecosystem through its API. Six years after its launch, it provides a large-scale global source of location data that describes real world places and human mobility in urban environments. 

In this paper we build on the legacy of Foursquare location data to provide new perspectives on urban activity growth patterns in cities. 
Working on a longitudinal multi-city snapshot of the data, we are interested in profiling cities based on their activity growth patterns. We first develop a methodology to track urban growth patterns, in terms of new place creation in cities.
Next, we propose a technique that exploits dynamic location data to detect areas in a city that surge in terms of development.
Finally, we investigate how the emergence of new urban activities in a neighbourhood influences existing locations in terms of foot traffic.
In detail, we make the following
contributions: 

\begin{itemize}

\item Initially we provide an inter-city perspective on urban activity growth by representing cities as vectors of the frequency of new place types within them. 
By applying spectral clustering to these vectors, we demonstrate a strong correlation between urban growth patterns and location, with nearby pairs of cities, such as those that belong to the same country or continent, being far more likely to share a similar urban activity profile than remote pairs of cities.
We hypothesise that these observations are a result of archetypal patterns of urban growth shared on a regional level and possibly rooted in similar cultures.
In fact, when we represent cities in terms of relative growth places (frequency of new place types versus frequency of existing place types), a method that aims to capture more recent policies affecting urban growth, we show that the spatial correlation in the clustering results is lost.

\item Next, we take a more local view on the urban growth process, performing a vertical analysis on each city.
Specifically, we analyse the spatial distribution of new places across the city, tracking \textit{where} new places tend to be created. 
While the influence of a strong urban hierarchy is prevalent with more new places being created in what is typically known as the urban core of a city, there are examples where accelerated growth in urban development occurs in peripheral areas.
Frequently this phenomenon is due to the existence of large development projects in response to preparation for large events such as the Olympic Games or the World Cup, as we demonstrate with representative case studies in London, UK and Bras\'{i}lia, Brazil.

\item Finally, we look at the impact of urban development on existing places. Exploiting user mobility information, we measure how the opening of a new venue can have an effect on local establishments in terms of foot traffic.
We identify the formation of two important trends: firstly, the existence of \textit{cooperative} place types that enable larger mobility flows to nearby venues, and secondly, the existence of \textit{competitive} place types whose presence in an area disrupts existing traffic flows to nearby places.
Interestingly, the former class of place types includes categories such as monuments, train stations or public spaces that represent anchors of generative urban development, whereas the latter category involves local businesses such as restaurants, pharmacies or barbershops that typically compete for customer traffic. 
There are exceptions however, a notable one being the presence of Turkish restaurants, which we discover tend to form local ecosystems that reinforce traffic volumes to other venues of the same type.

\end{itemize}

Overall, our analysis shows how modern datasets, generated by mobile users as they naturally explore an urban environment, can form the basis for sustainable monitoring frameworks and tools that could be deployed to manage tomorrow's cities.

\label{sec:analysis}
\section{The Dataset} \label{sec:dataset}

The foundation of our analysis is a four-year long dataset comprised of mobility records describing movements between places in 100 cities from around the globe. 
For each Foursquare venue in a city we know its geographic coordinates, its category or place type (Nightlife spot, Travel \& Transport, Hotel etc.), its creation time and the total number of users that have \textit{checked in} there.
For the purposes of this paper, we conceptually associate a place type, alternatively noted as a place category, to the \textit{urban activity} that is directly implied by it.
One of the criteria for a venue to have been included in the dataset is that it must have at least 100 total check-ins.
This hopefully removes the majority of false venues added by users, either mistakenly or maliciously.

Alongside the venue data, for each city the dataset also contains all ``transitions" occurring within the city.
A transition is defined to be a pair of check-ins by a single user to two different venues separated by less than 3 hours. For each transition we have the \textit{start time}, \textit{end time}, \textit{source venue} and \textit{destination venue}.
The transition records contain no information about the identity of the user. 

Critically we have information on the creation time of a place (i.e. the time that the venue was added to the Foursquare database).
In Appendix~\ref{app:new_places}, we provide a detailed description of the filtering methodology we apply to discriminate newly created venues from those that had already existed and were crowd-sourced to the venue database by mobile users at a time later than the actual opening time. 

\section{Macro-scale analysis}

\subsection{City growth profiles} \label{sec:city_clustering}
In this section we demonstrate that data crowdsourced from location-based services can be used to identify cities and regions where particular  urban activities are currently experiencing strong growth.
To begin with we investigate clustering the cities according to their growth profile in terms of the new activities that emerge in the urban territory.
The main questions we address are: \textit{Can we exploit crowdsourced data about places to profile
cities in terms of urban activity growth?} and moreover
\textit{How do these profiles of urban growth compare and what is the role of geography in this relationship?}

These questions naturally arise from the observation that cities are famous for their particular composition of place types. For example many buildings in Cambridge, UK are directly or indirectly related to the local university and Paris is well known for its many caf\'{e}s.
In this work, we put forward a methodology where   the focus is not just on the overall compositions of place types in cities, but how their composition is currently changing as new places are being created.  
While the nature of such change is interesting in of itself, as it can shed light on historic and cultural aspects of urban growth, it can also be viewed as a snapshot of current investment in the city.
This can highlight the priorities of local government and public spending (Colleges \& Universities, Schools, Government Buildings etc.) as well as where growth in the private sector is currently focused (Hotels, Food, Offices etc.). 

\subsubsection*{Activity growth vectors}
For each city we generate a growth vector $\vec{v}$ whose $i^{th}$ component, $\vec{v}_i$, represents the proportion of new venues in general category $i$ as calculated in Equation~\ref{eq:growth_vector}.
\begin{equation} \label{eq:growth_vector}
\vec{v}_i = \frac{N_{new}(i)}{\sum_i N_{new}(i)}
\end{equation}
The value $N_{new}(i)$ is the number of new places in the city in category $i$. 

\begin{figure}[ht]

	\newcommand{\cgpwidth}[0]{0.9\columnwidth}

    \centering
	\begin{subfigure}{\cgpwidth}
	    \includegraphics[width=\columnwidth]{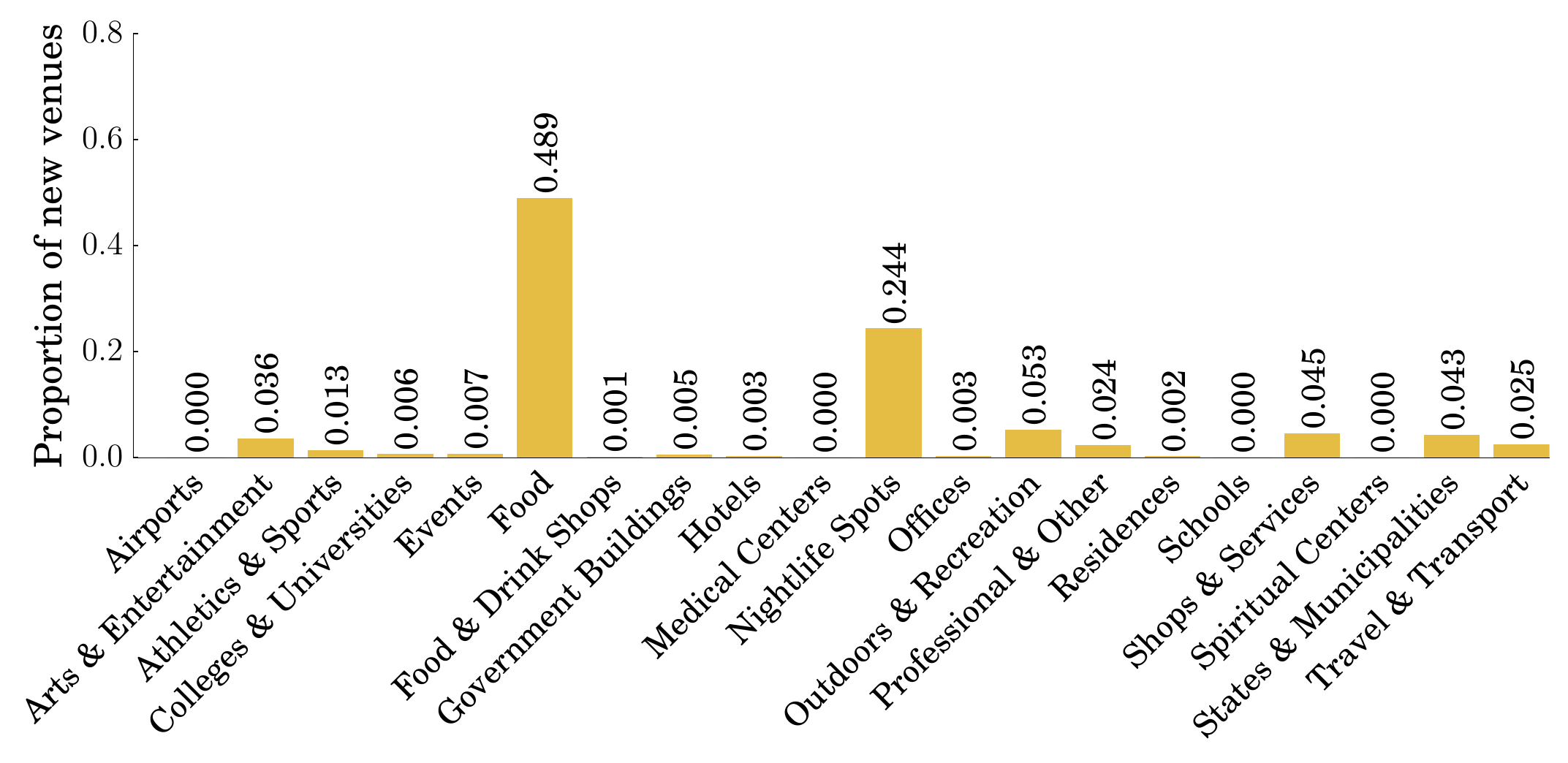}
	    \caption{Athens}
	\end{subfigure}
	
	\begin{subfigure}{\cgpwidth}
	   	\includegraphics[width=\columnwidth]{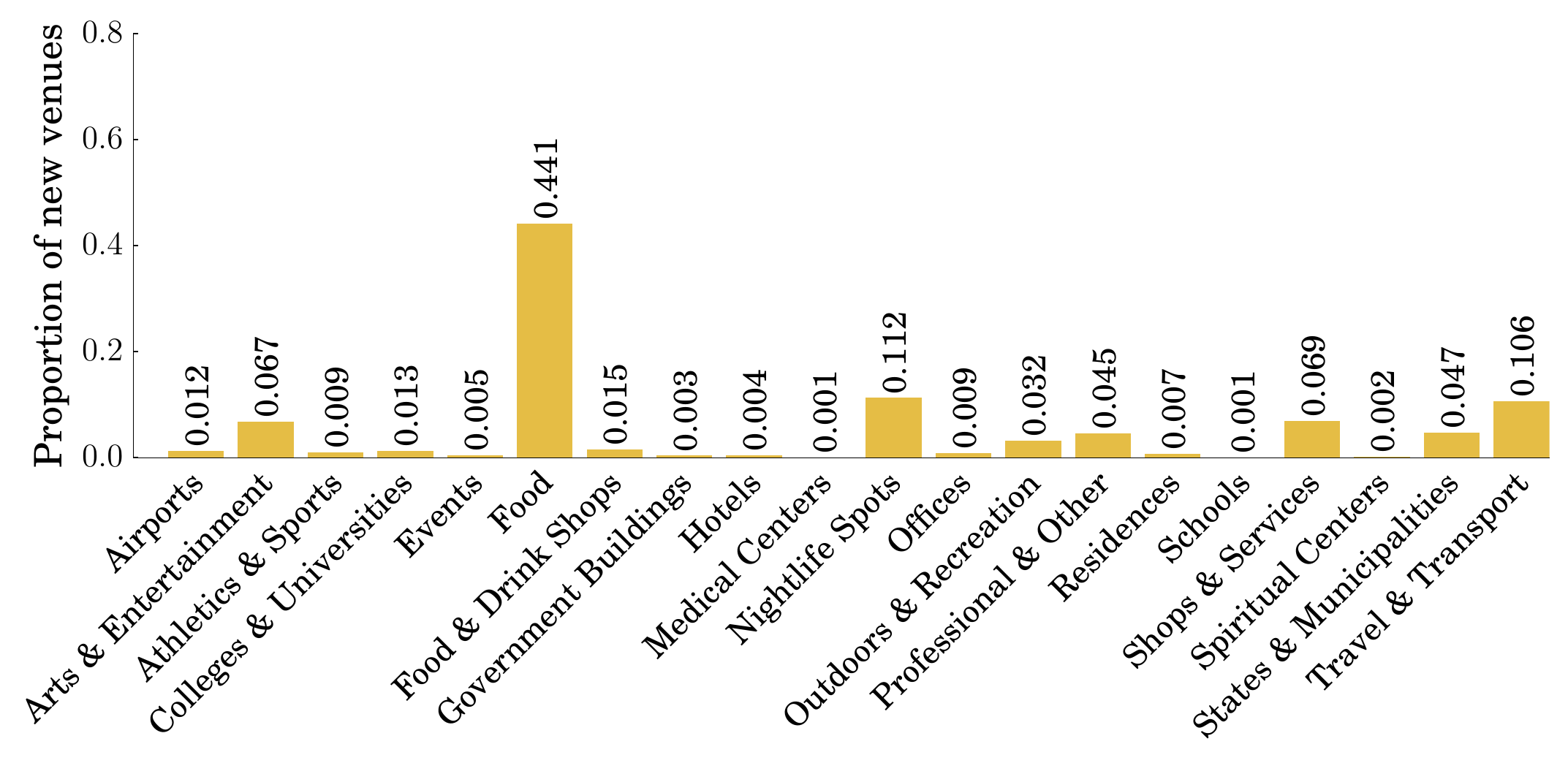}
	    \caption{London}
	\end{subfigure}
	    	
	    	
    \begin{subfigure}{\cgpwidth}
    	\includegraphics[width=\columnwidth]{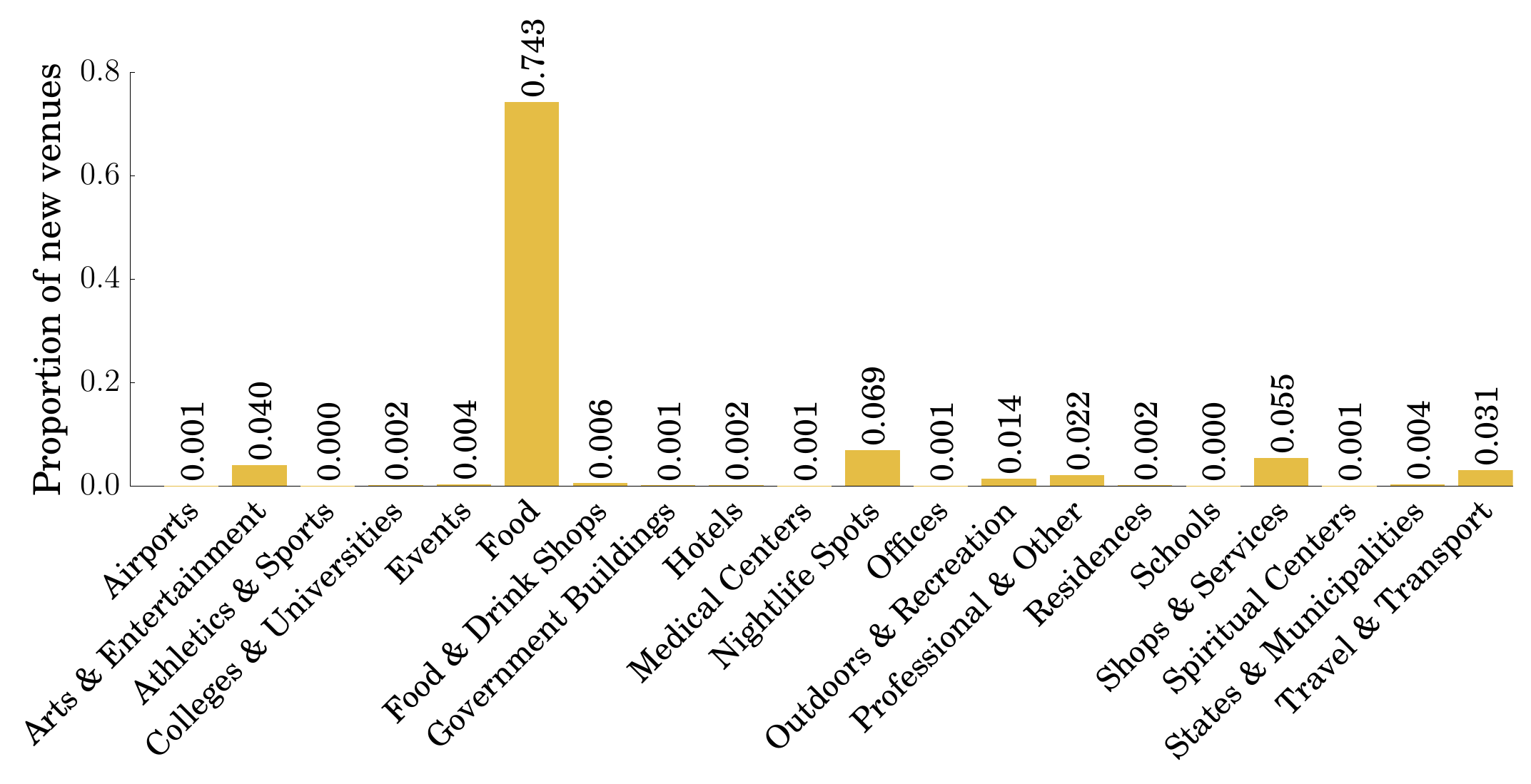}
    	\caption{Seoul}
    \end{subfigure}
    
    \begin{subfigure}{\cgpwidth}
    	\includegraphics[width=\columnwidth]{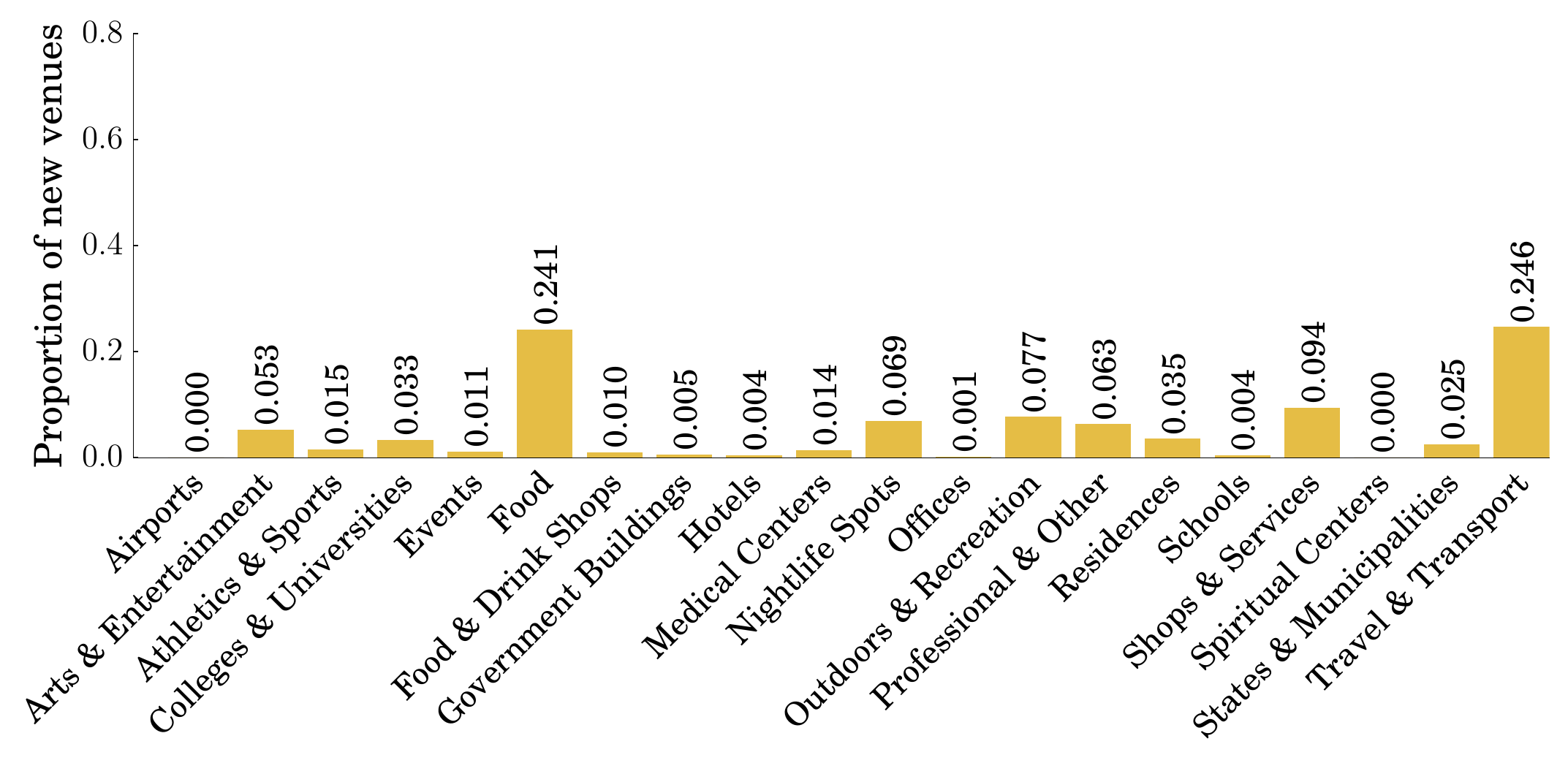}
    	\caption{Riga}
    \end{subfigure}
    \caption{Activity growth profiles for four different cities}
    \label{fig:category_growth_profiles}
\end{figure}

As an example, the growth vectors for Seoul, Athens, Riga and London, are shown in Figure~\ref{fig:category_growth_profiles}.
The Food category is shown to contribute a significant proportion of all new place growth, ranging from 24\% in Riga to 74\% in Seoul. This is to be expected due to the entrepreneurial nature of food establishments. Accordingly the average lifespan of Food venues is relatively low, and so the category undergoes significant churn. 
Despite the prominence of Food establishments that is common across cities, we can also observe differences in the dynamics of urban activity profiles of cities.
London and Riga tend to grow with much higher rates regarding their transport infrastructure (Travel \& Transport places), whereas Athens shows a blossoming nightlife scene amidst the Greek financial crisis.

\subsubsection*{Clustering Methodology}
Louf et \emph{al.} \cite{louf14} apply a methodology for clustering cities 
based on the topology of street networks and the geometry of the land patches
that emerge amongst the networks of urban streets.
They find groups of cities that present similar patterns and trace these observations to different urban planning and design strategies adopted by urban authorities. 
Here, we use a different layer of information that describes a city, specifically the place types or urban activities that emerge as new places open. 
We cluster cities based on the urban activity growth profiles described in the previous paragraph. 
Compared to Louf et \emph{al.} we use spectral~\cite{louf14}, rather than hierarchical, clustering, as it does not require an initial assumption about the number of clusters \cite{ng02}.
Silva et \emph{al.}~\cite{Silva14} also apply a methodology based on spectral clustering on Foursquare data to measure cultural similarities between cities in terms of Food establishments and culinary preferences. 

\newcommand{\csize}{1.0em}
\begin{figure}[!ht]
	\centering
	\includegraphics[width=\columnwidth]{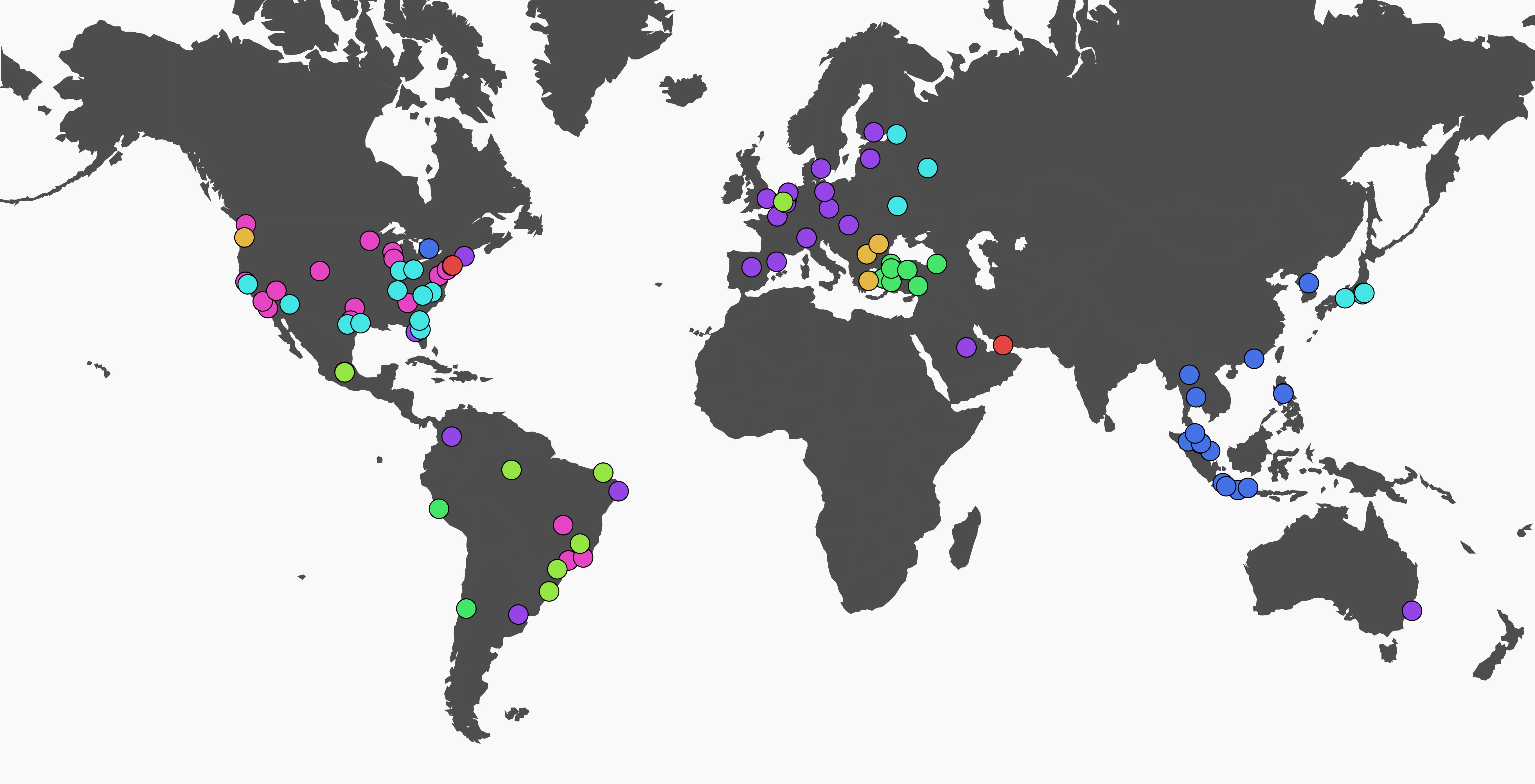}
	
	\vspace{1em}

	\begin{tabularx}{\columnwidth}{ccX}
	\toprule
	\# & Colour & Cities \\
	\midrule
	1 & \includegraphics[width=\csize] {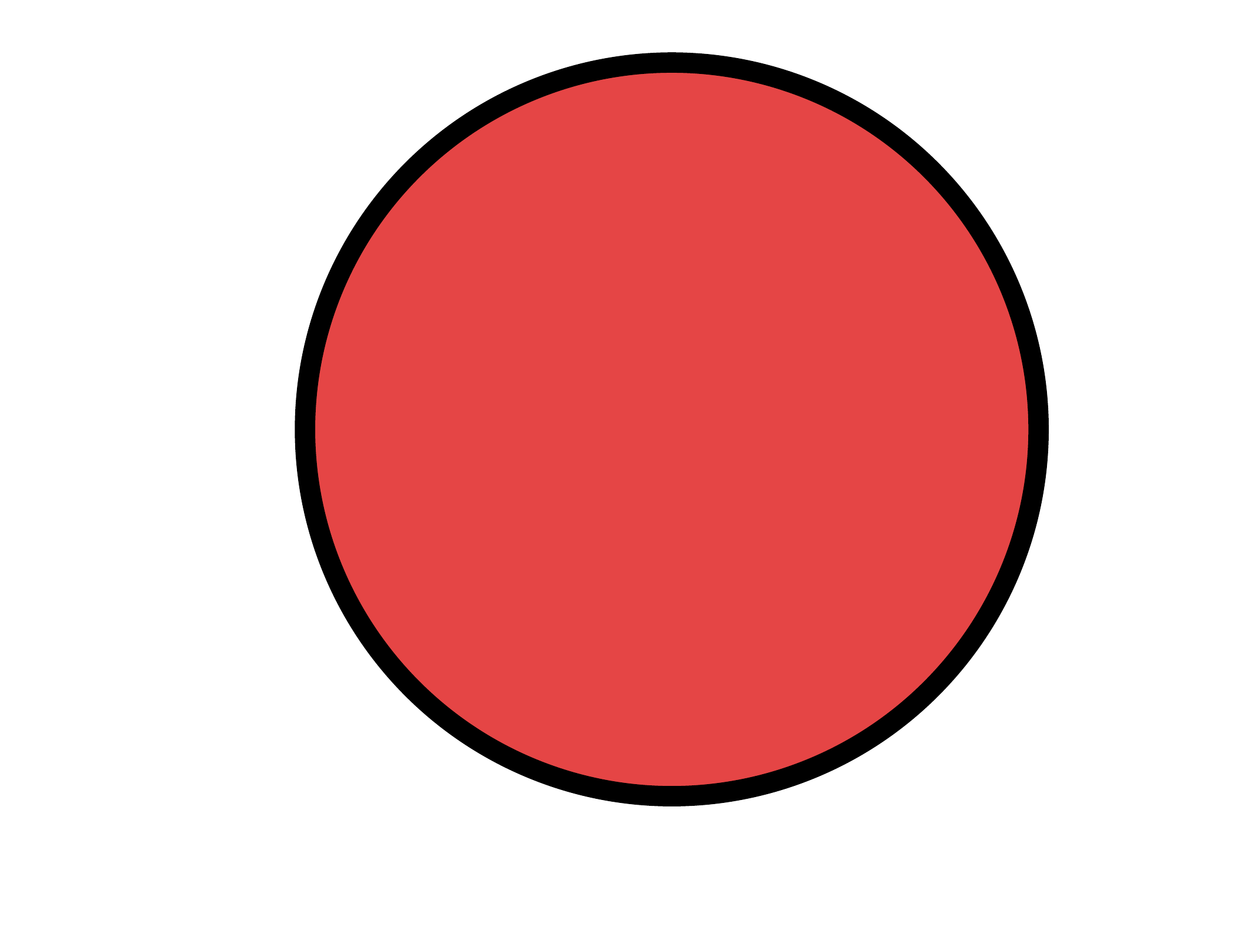} & Dubai, Borough of Queens \\
	2 & \includegraphics[width=\csize]{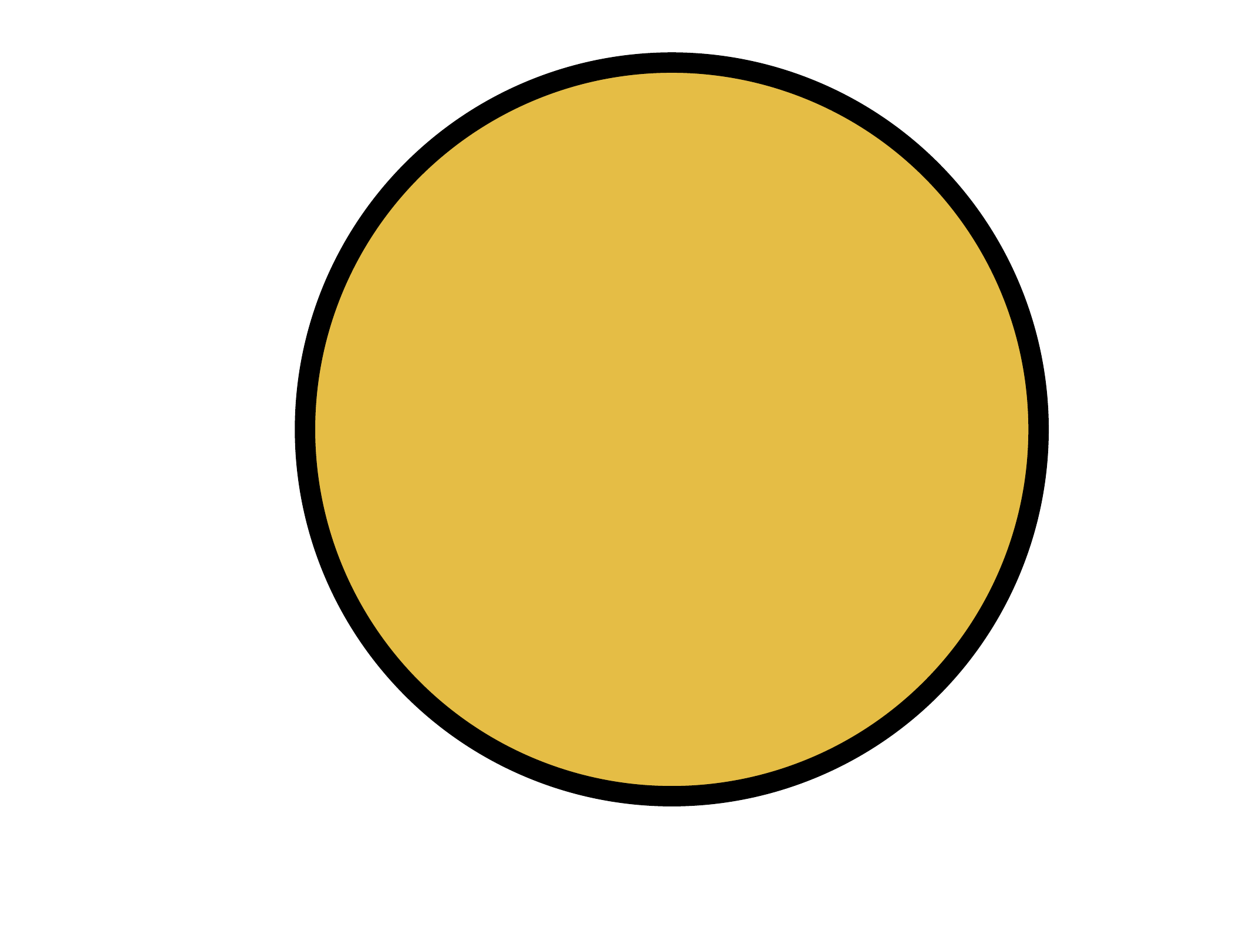} & Athens, Brooklyn, Bucharest, Portland, Sofia \\
	3 & \includegraphics[width=\csize]{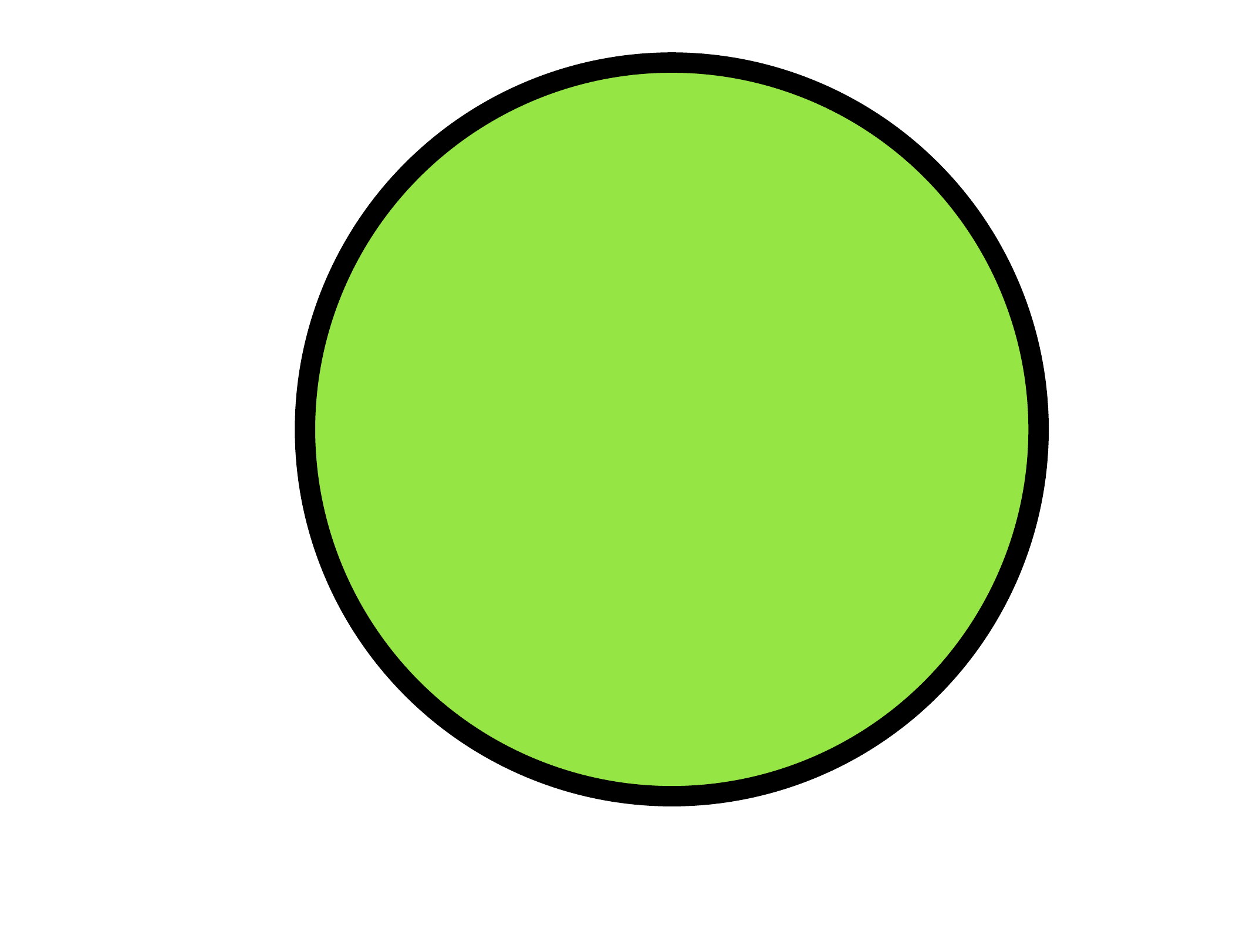} & Belo Horizonte, Coyoac\`{a}n, Curitiba, Fortaleza, Gent, Manaus, Porto Alegre \\
	4 & \includegraphics[width=\csize]{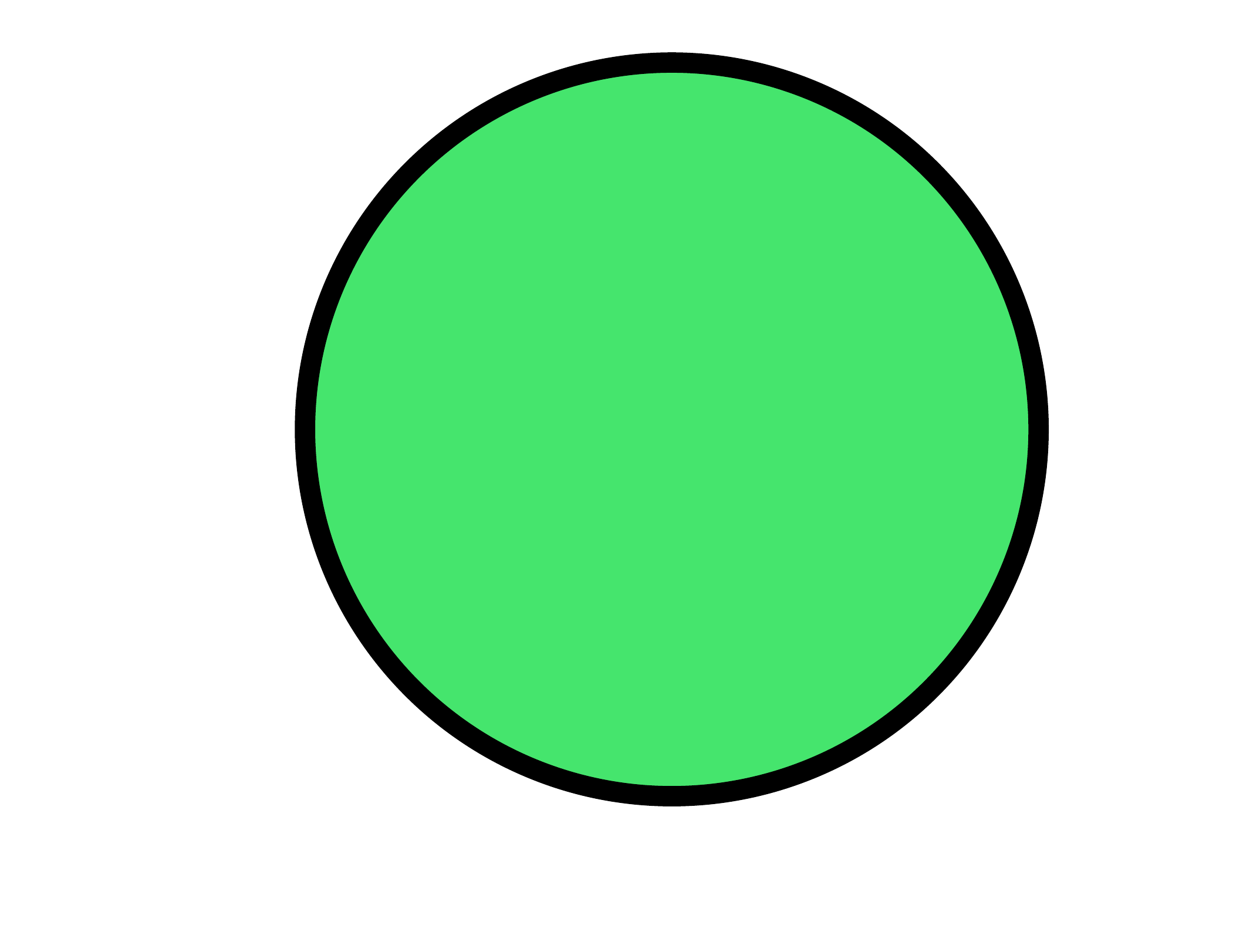} & Adana, Ankara, Bursa, Denizli, Eski\c{s}ehir, \.{I}stanbul, \.{I}zmir, Lima, Santiago, Trabzon \\
	5 & \includegraphics[width=\csize]{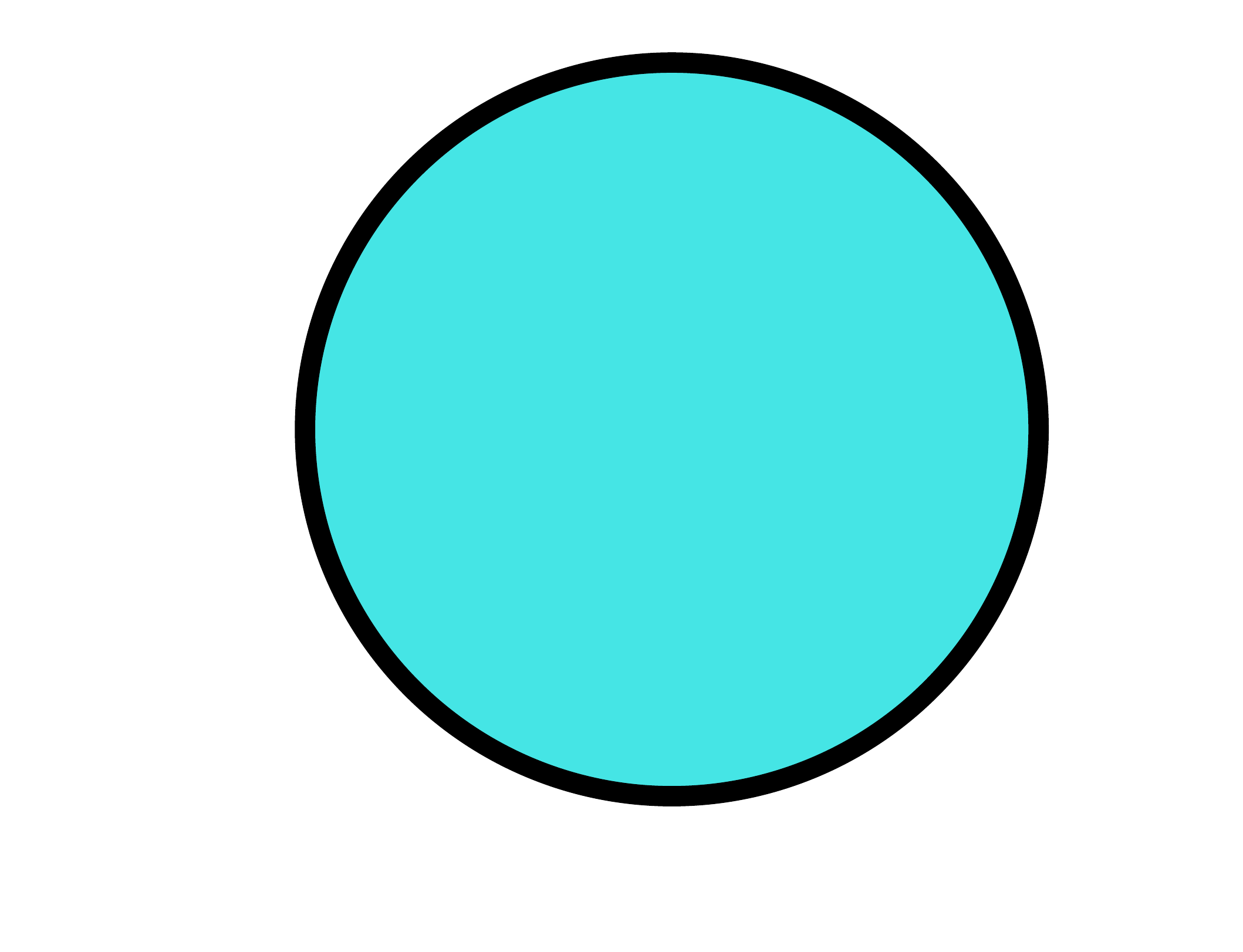} & Charlotte, Chiba, Columbus, Houston, Indianapolis, Jacksonville, Kiev, Moscow, Nashville, Orlando, Osaka, Phoenix, Raleigh, Saint Petersburg, San Antonio, San Jose, Yokohama \\
	6 & \includegraphics[width=\csize]{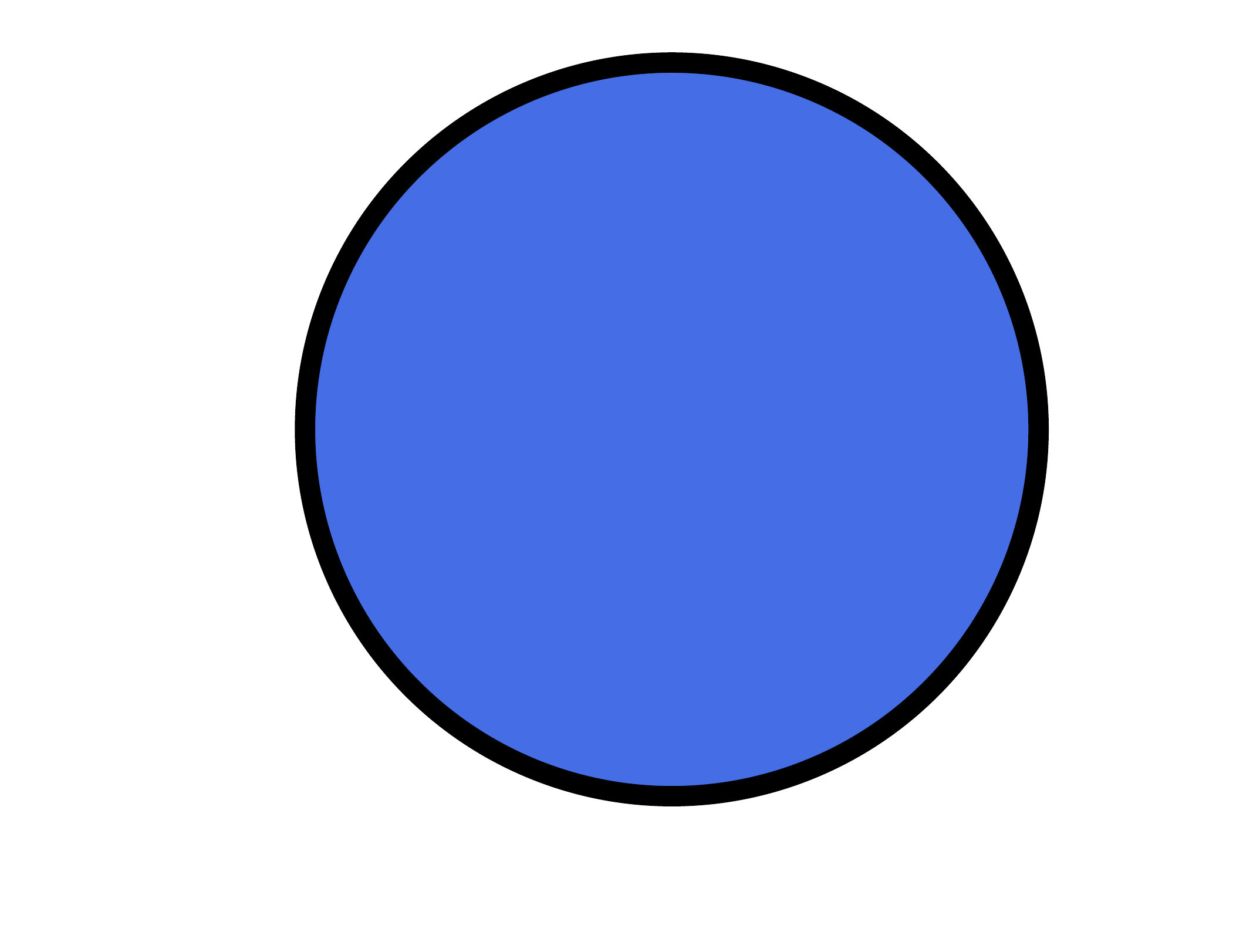} & Bandung, Bangkok, Chiang Mai, George Town, Hong Kong, Jakarta, Kuala Lumpur, Makati City, Medan, Petaling Jaya, Pineda, Quezon City, Seoul, Shah Alam, Singapore, Surabaya, Tokyo, Toronto, Yogyakarta\\
	7 & \includegraphics[width=\csize]{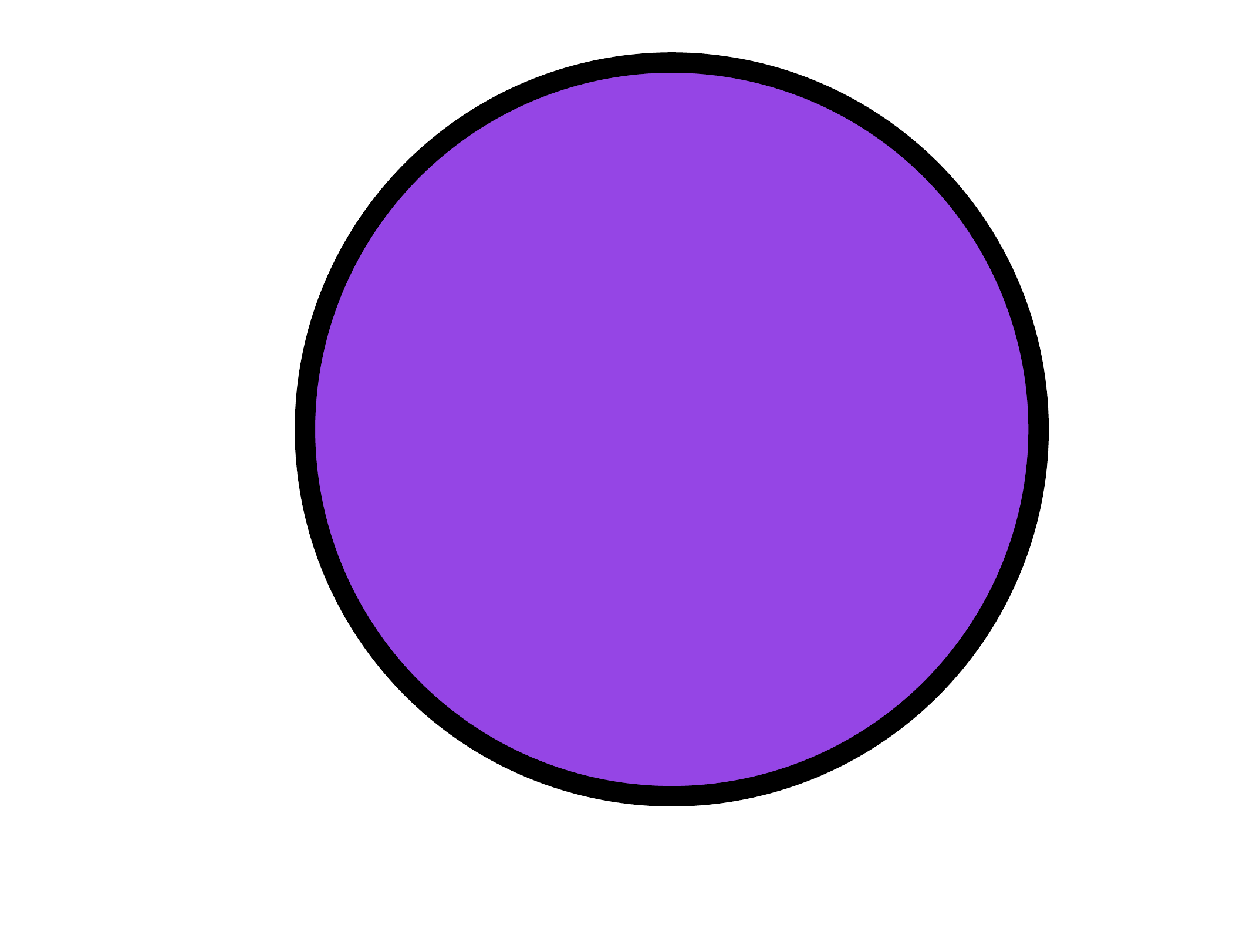} & Amsterdam, Barcelona, Berlin, Bogot\`{a}, Boston, Brussels, Budapest, Buenos Aires, Copenhagen, Helsinki, London, Madrid, Milano, Paris, Prague, Recife, Riga, Riyadh, Sydney, Tampa \\
	8 & \includegraphics[width=\csize]{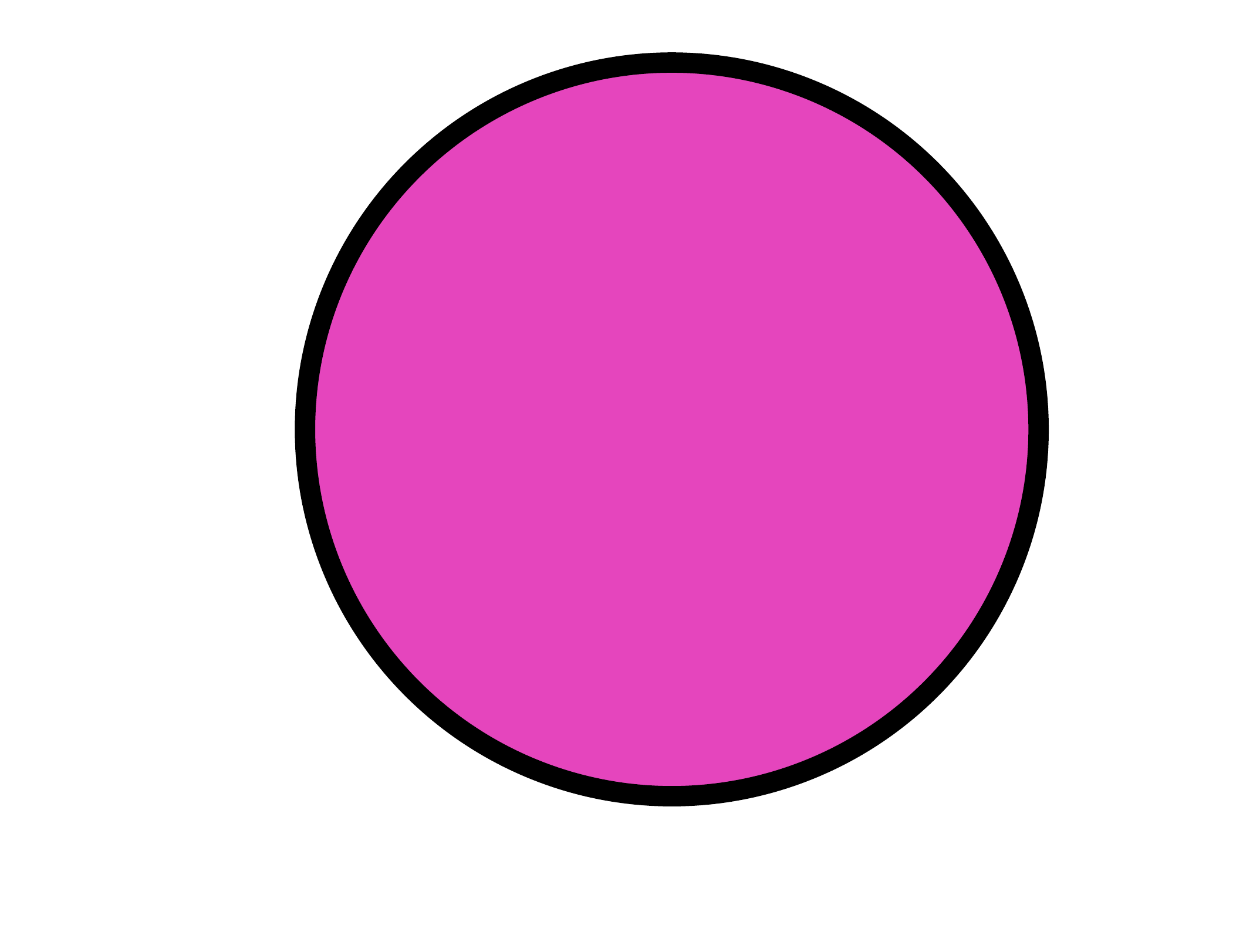} & Antwerpen, Atlanta, Austin, Bras\'{i}lia, Chicago, Dallas, Denver, Las Vegas, Los Angeles, Mexico City, Milwaukee, Minneapolis, New York, Philadelphia, Rio de Janeiro, San Diego, San Francisco, S\~{a}o Paulo, Seattle, Washington D.C. \\
	\bottomrule
	\end{tabularx}
	\caption{New growth vector clustering results}
	\label{fig:clustering_map}
\end{figure}

\begin{figure}
	\centering
	\captionsetup{justification=centering}
	\begin{tabular}{cccc}
 		\begin{subfigure}{0.2\columnwidth}
			\includegraphics[width=\columnwidth]{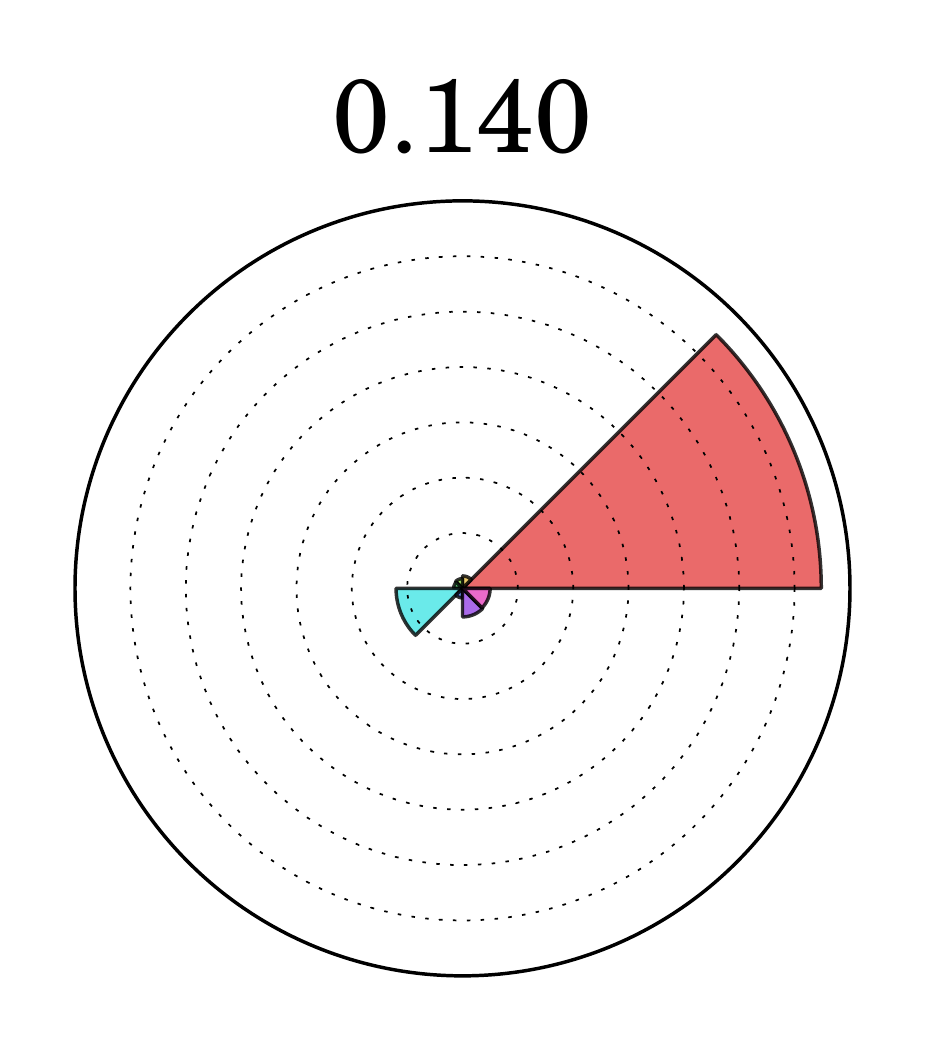}
			\caption*{\hfill Airports \hfill \break}
		\end{subfigure} &
		
		\begin{subfigure}{0.2\columnwidth}
			\includegraphics[width=\columnwidth]{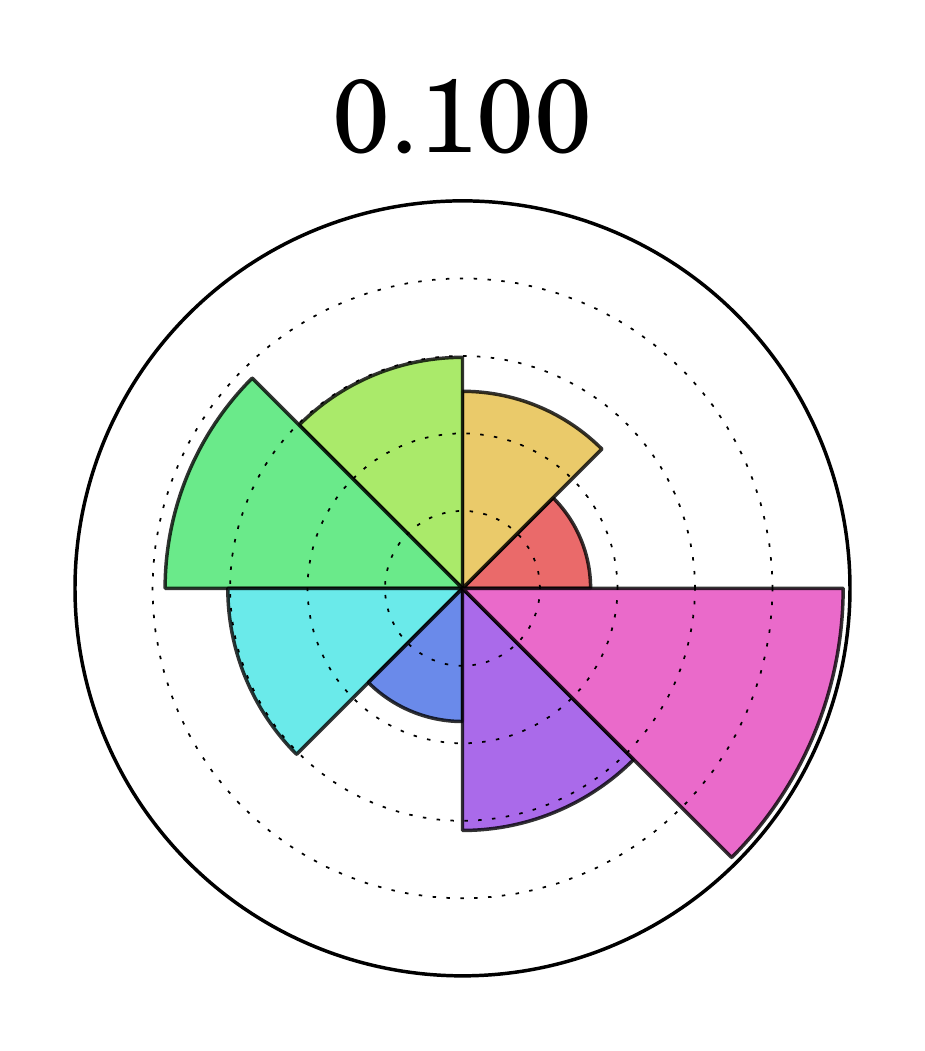}
			\caption*{Arts \& Entertainment}
		\end{subfigure} &
		
		\begin{subfigure}{0.2\columnwidth}
			\includegraphics[width=\columnwidth]{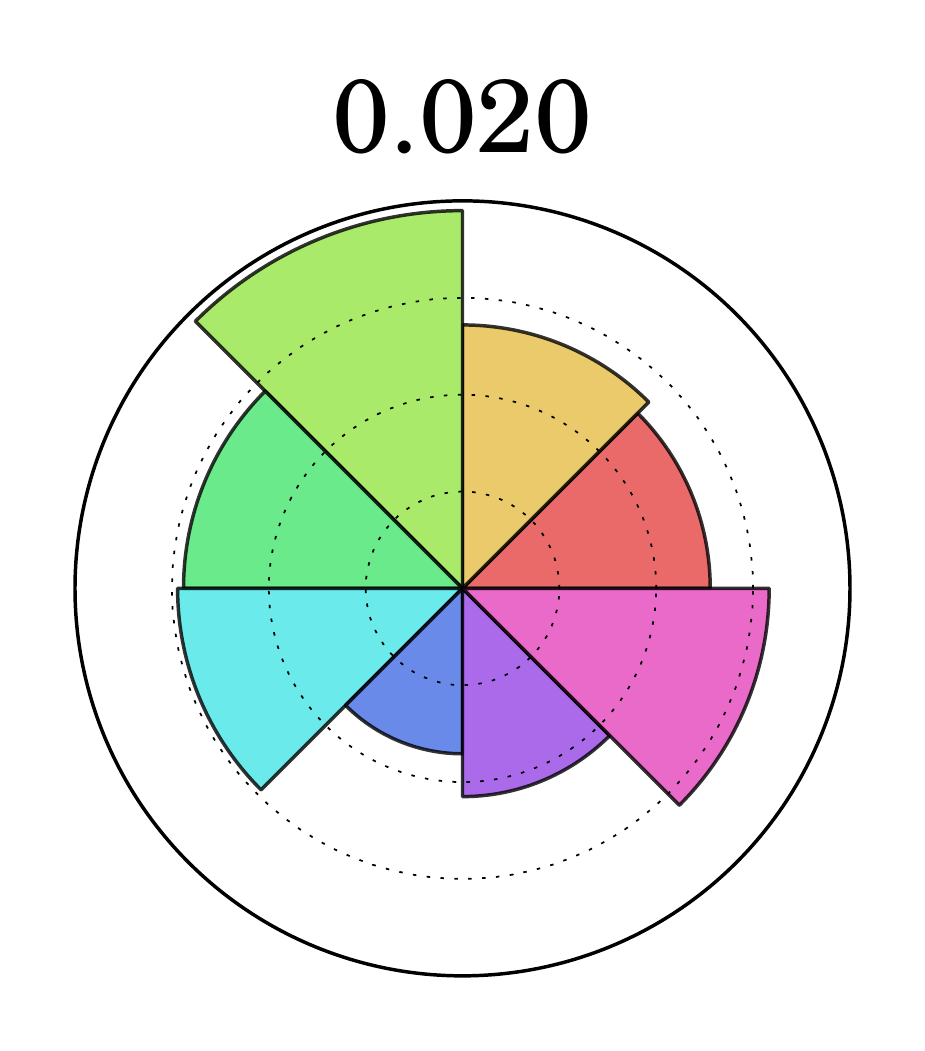}
			\caption*{Athletics \& Sports}
		\end{subfigure} &
		
		\begin{subfigure}{0.2\columnwidth}
			\includegraphics[width=\columnwidth]{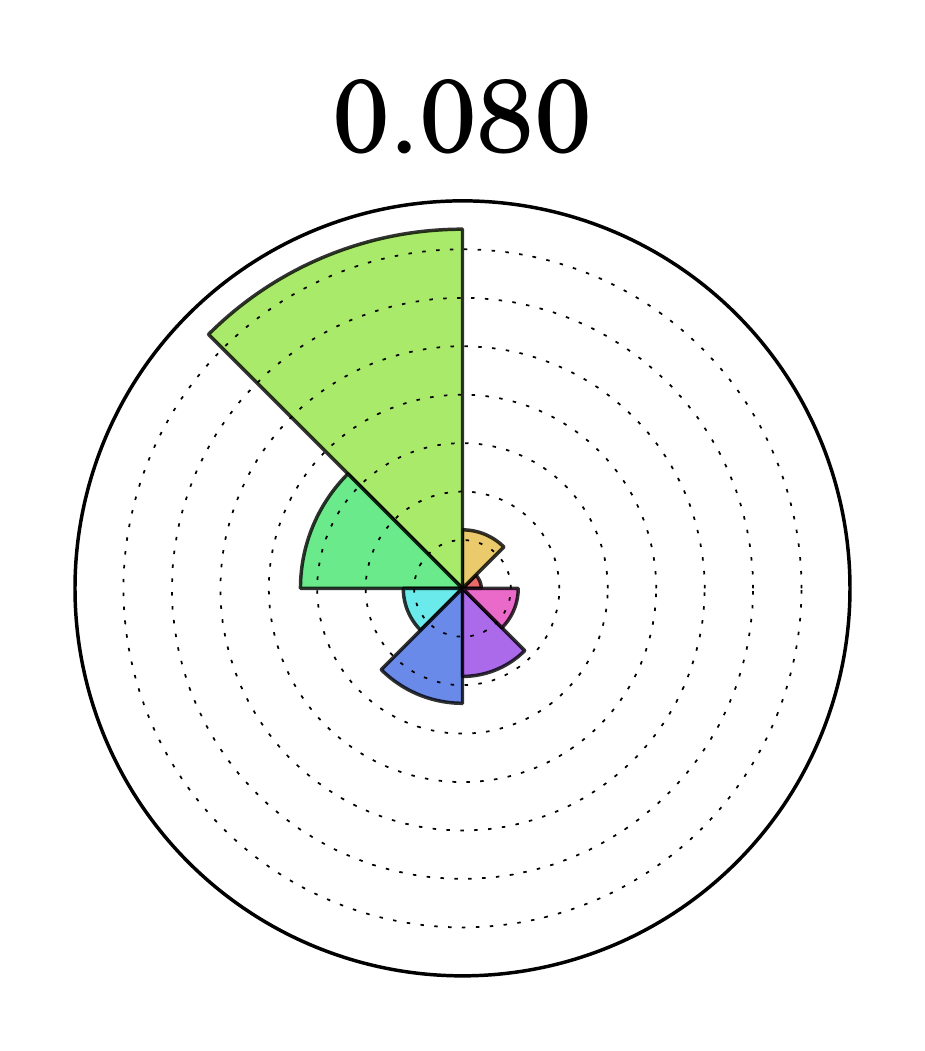}
			\caption*{Colleges \& Universities}
		\end{subfigure} \\

		\begin{subfigure}{0.2\columnwidth}
			\includegraphics[width=\columnwidth]{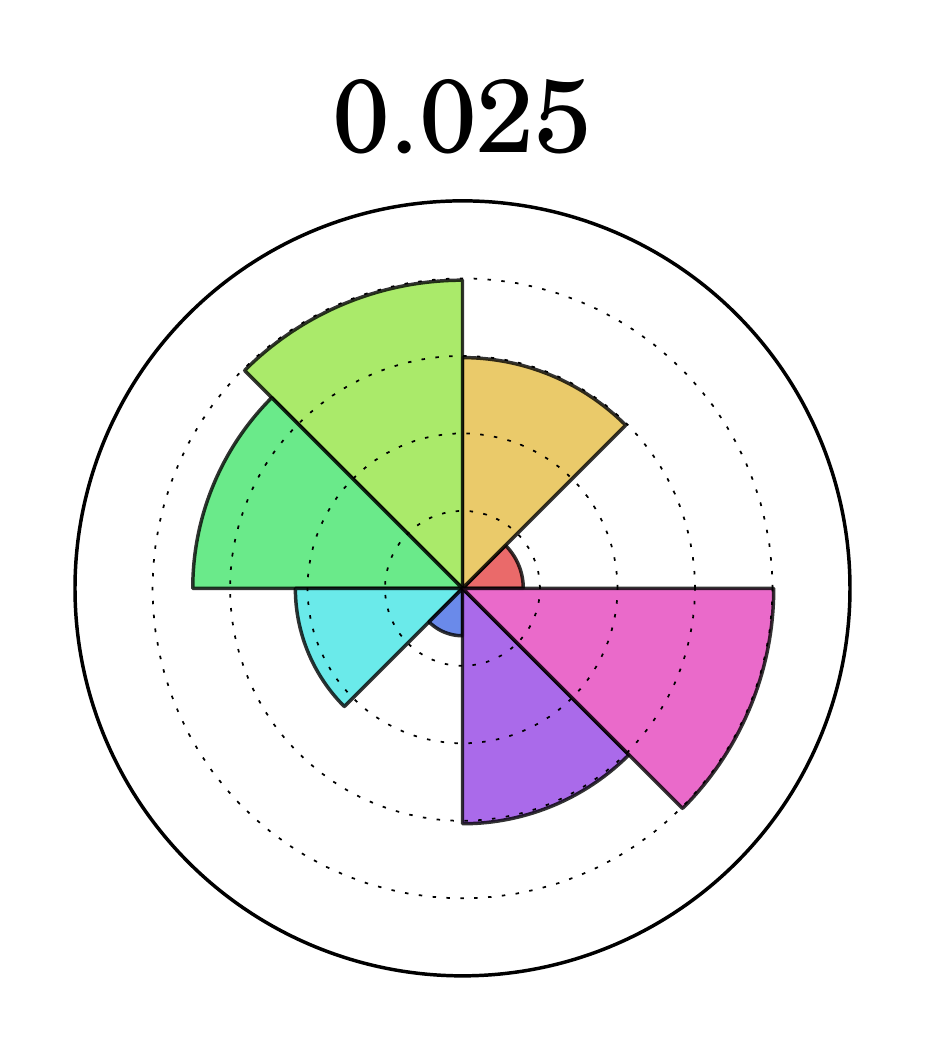}
			\caption*{\hfill Events \hfill \break}
		\end{subfigure} &
		
		\begin{subfigure}{0.2\columnwidth}
			\includegraphics[width=\columnwidth]{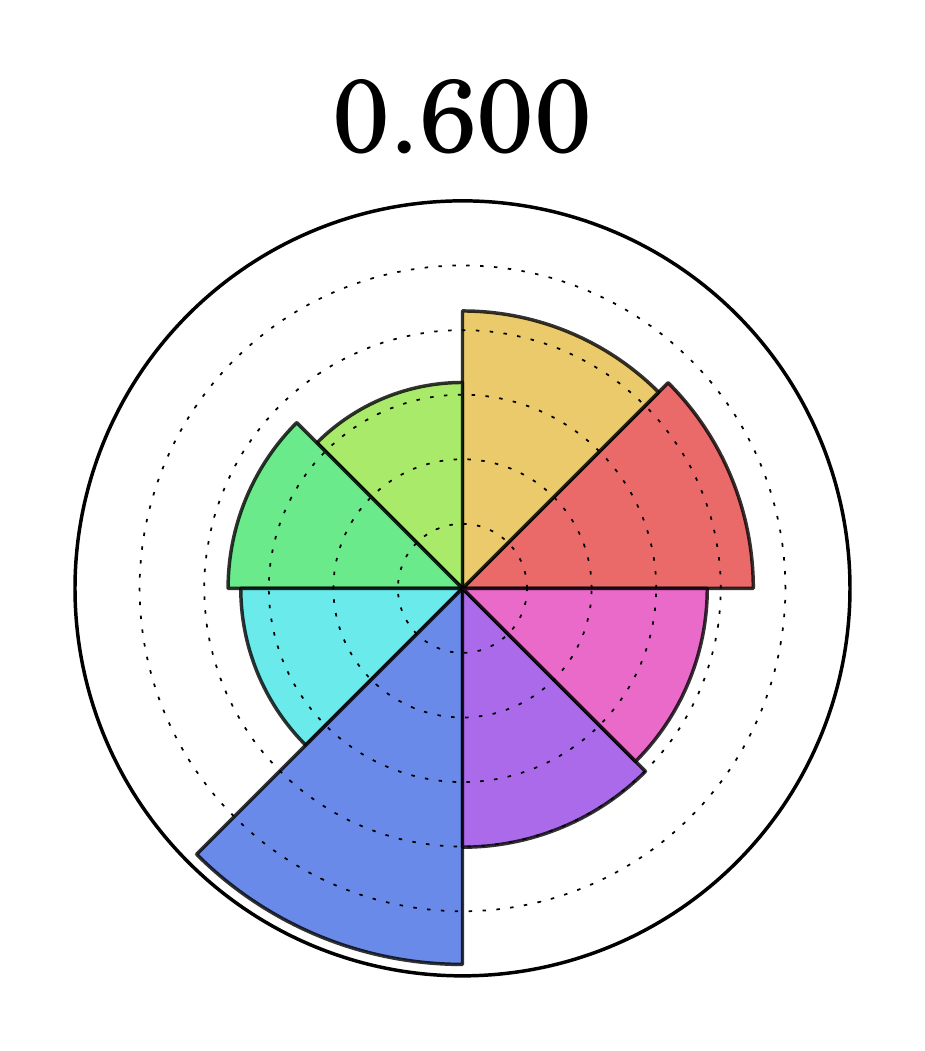}
			\caption*{\hfill Food \hfill \break}
		\end{subfigure} &
		
		\begin{subfigure}{0.2\columnwidth}
			\includegraphics[width=\columnwidth]{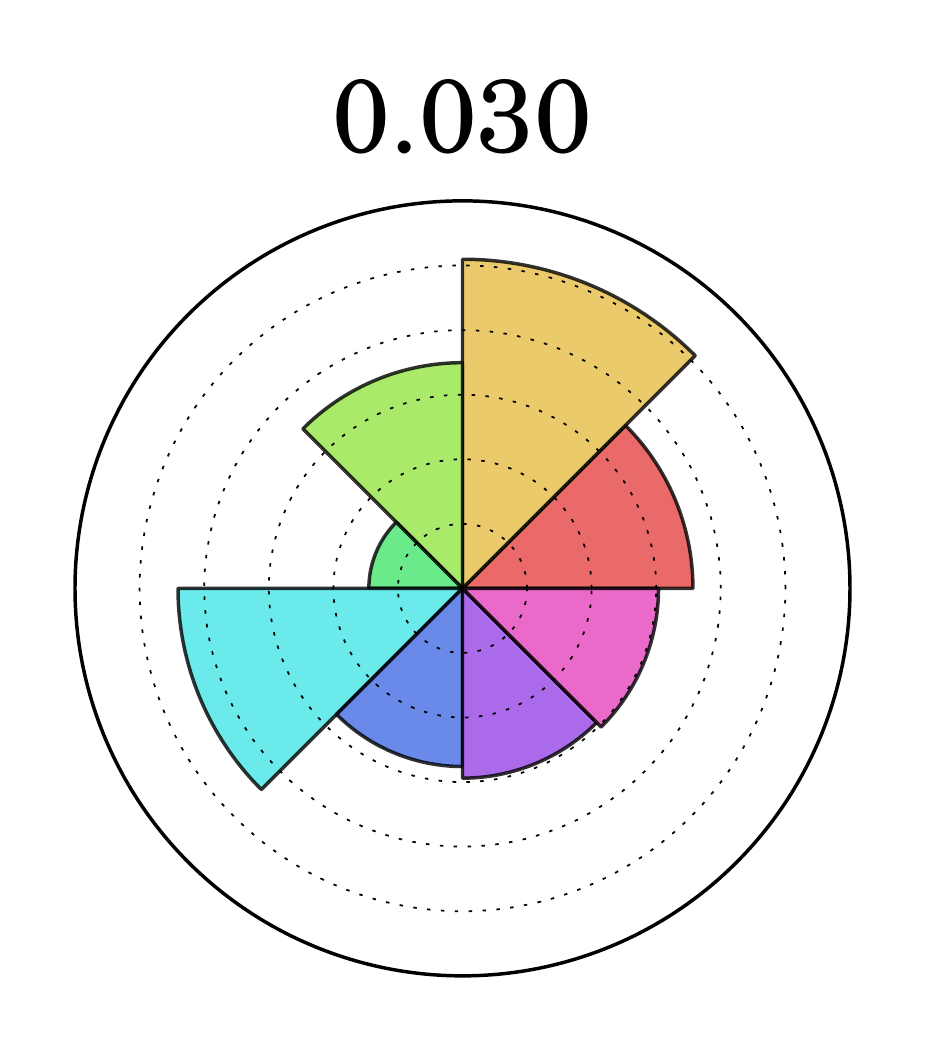}
			\caption*{Food \& Drink Shops}
		\end{subfigure} &
		
		\begin{subfigure}{0.2\columnwidth}
			\includegraphics[width=\columnwidth]{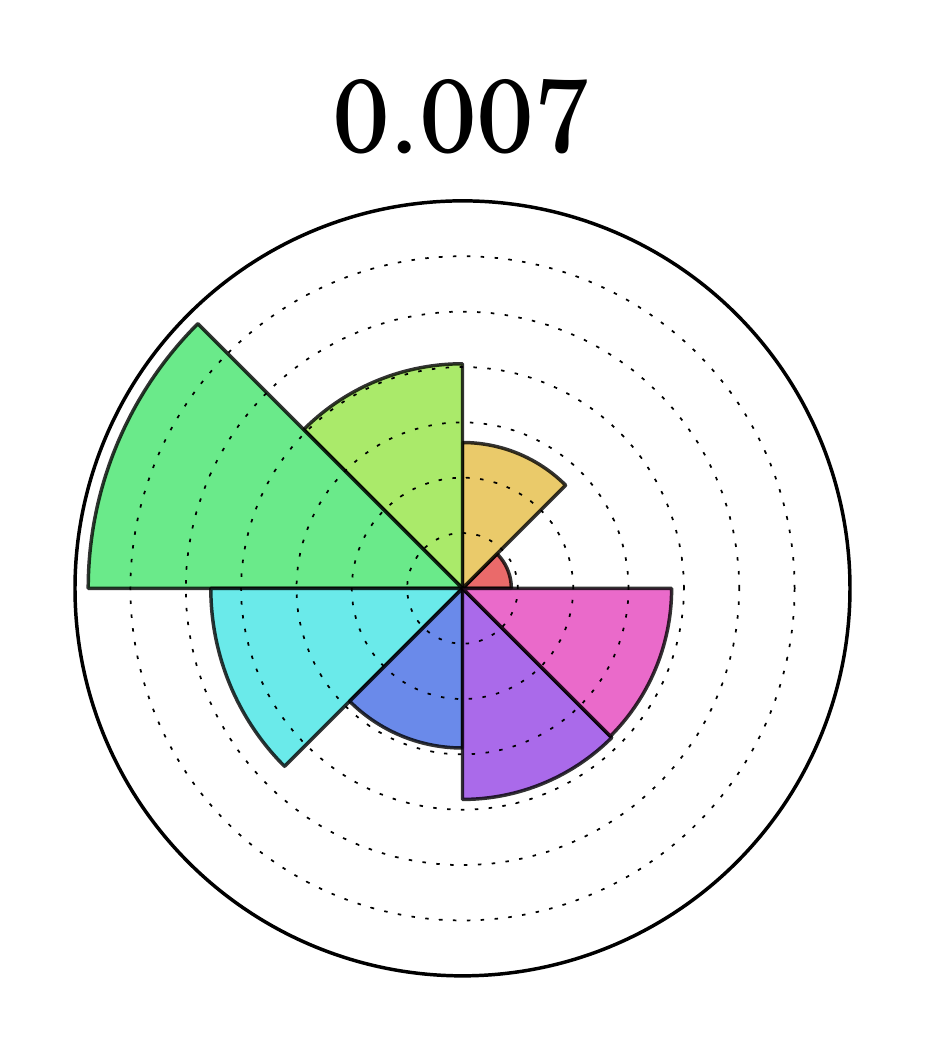}
			\caption*{Government Buildings}
		\end{subfigure} \\

		\begin{subfigure}{0.2\columnwidth}
			\includegraphics[width=\columnwidth]{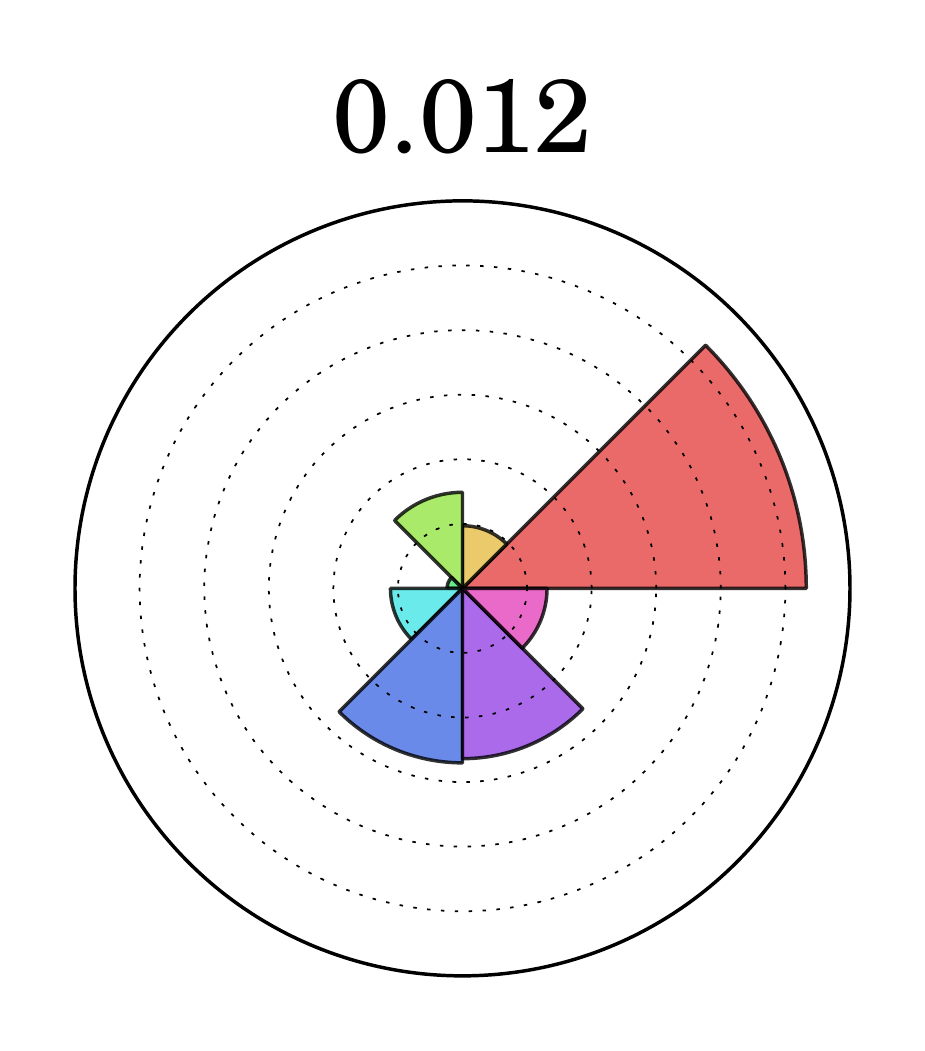}
			\caption*{\hfill Hotels \hfill \break}
		\end{subfigure} &
		
		\begin{subfigure}{0.2\columnwidth}
			\includegraphics[width=\columnwidth]{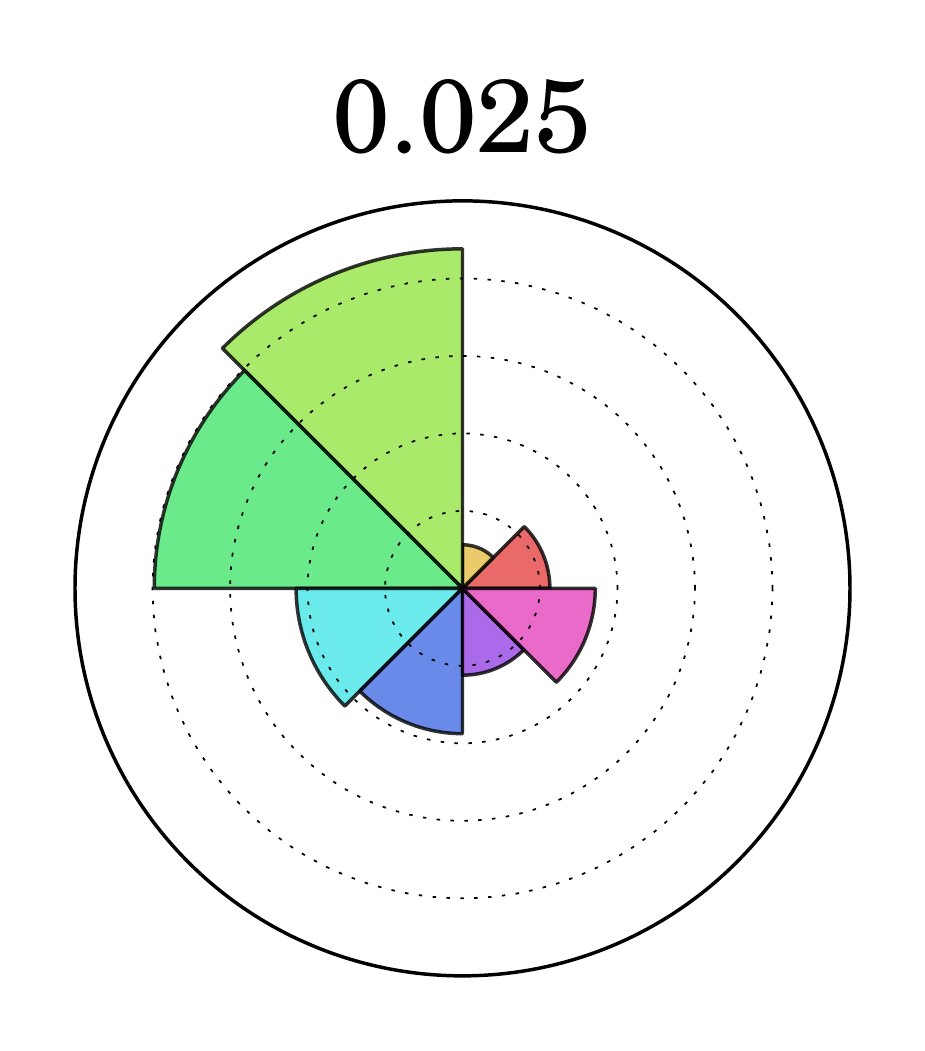}
			\caption*{Medical Centers}
		\end{subfigure} &
		
		\begin{subfigure}{0.2\columnwidth}
			\includegraphics[width=\columnwidth]{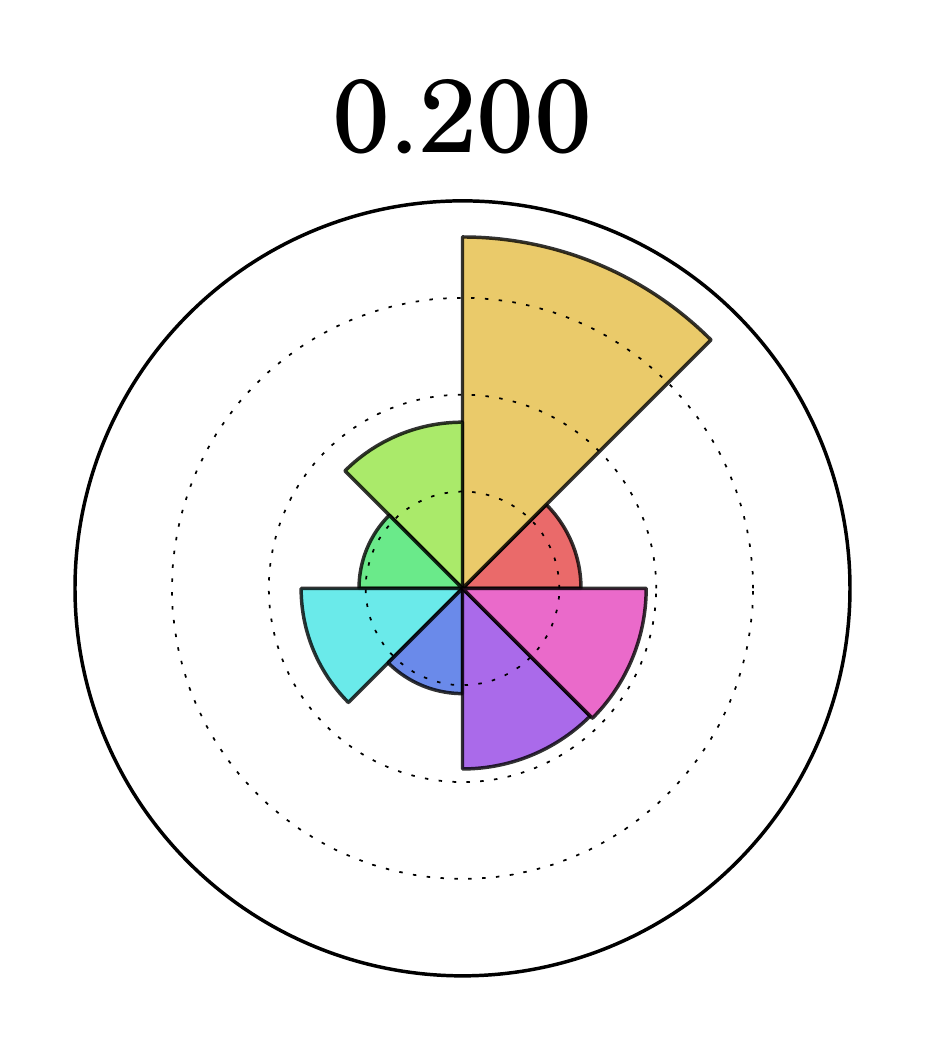}
			\caption*{Nightlife Spots}
		\end{subfigure} &
		
		\begin{subfigure}{0.2\columnwidth}
			\includegraphics[width=\columnwidth]{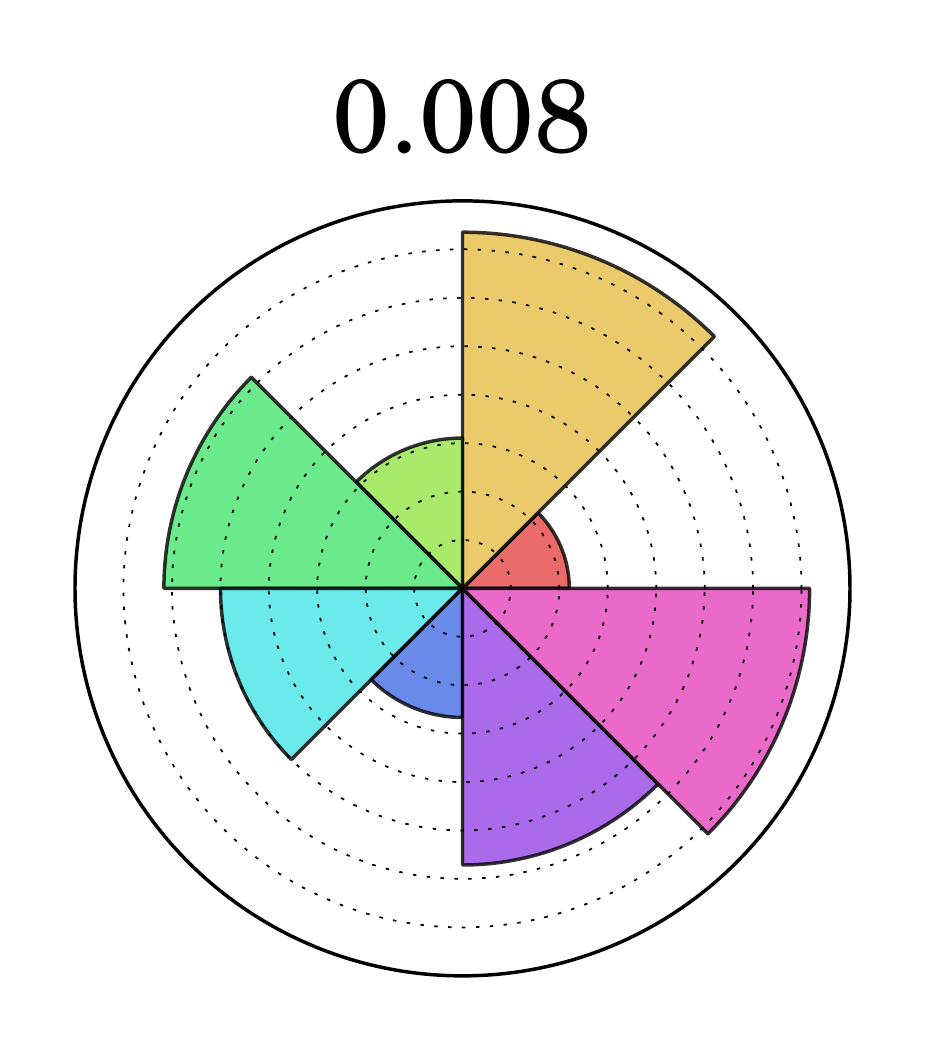}
			\caption*{\hfill Offices \hfill \break}
		\end{subfigure} \\

		\begin{subfigure}{0.2\columnwidth}
			\includegraphics[width=\columnwidth]{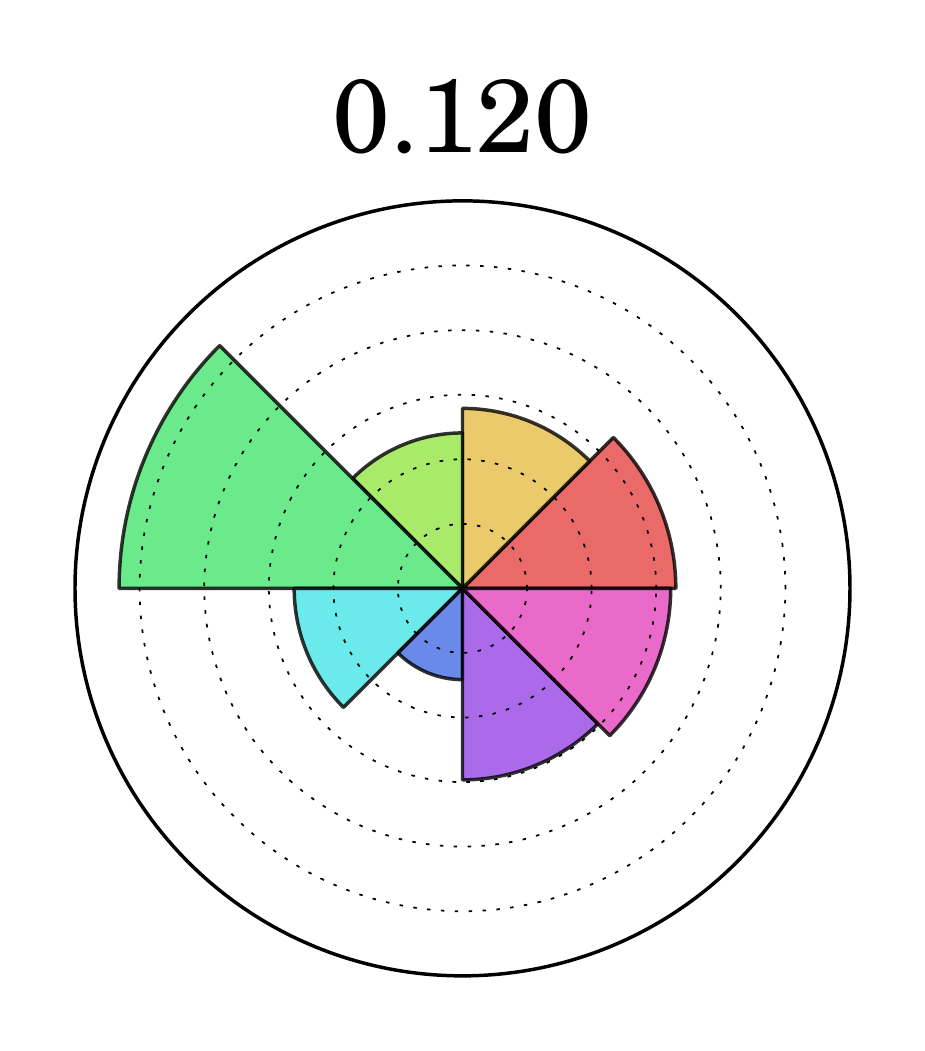}
			\caption*{Outdoors \& Recreation}
		\end{subfigure} &
		
		\begin{subfigure}{0.2\columnwidth}
			\includegraphics[width=\columnwidth]{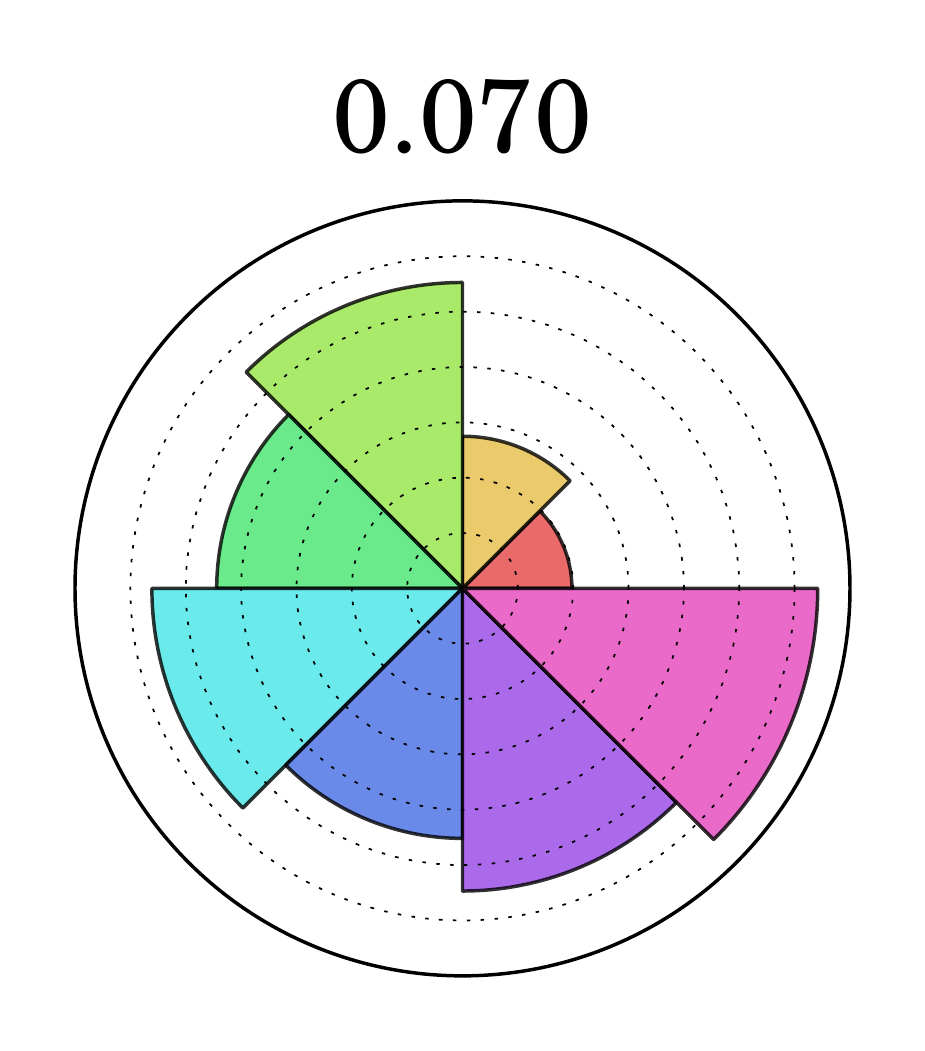}
			\caption*{Professional \& Other}
		\end{subfigure} &
		
		\begin{subfigure}{0.2\columnwidth}
			\includegraphics[width=\columnwidth]{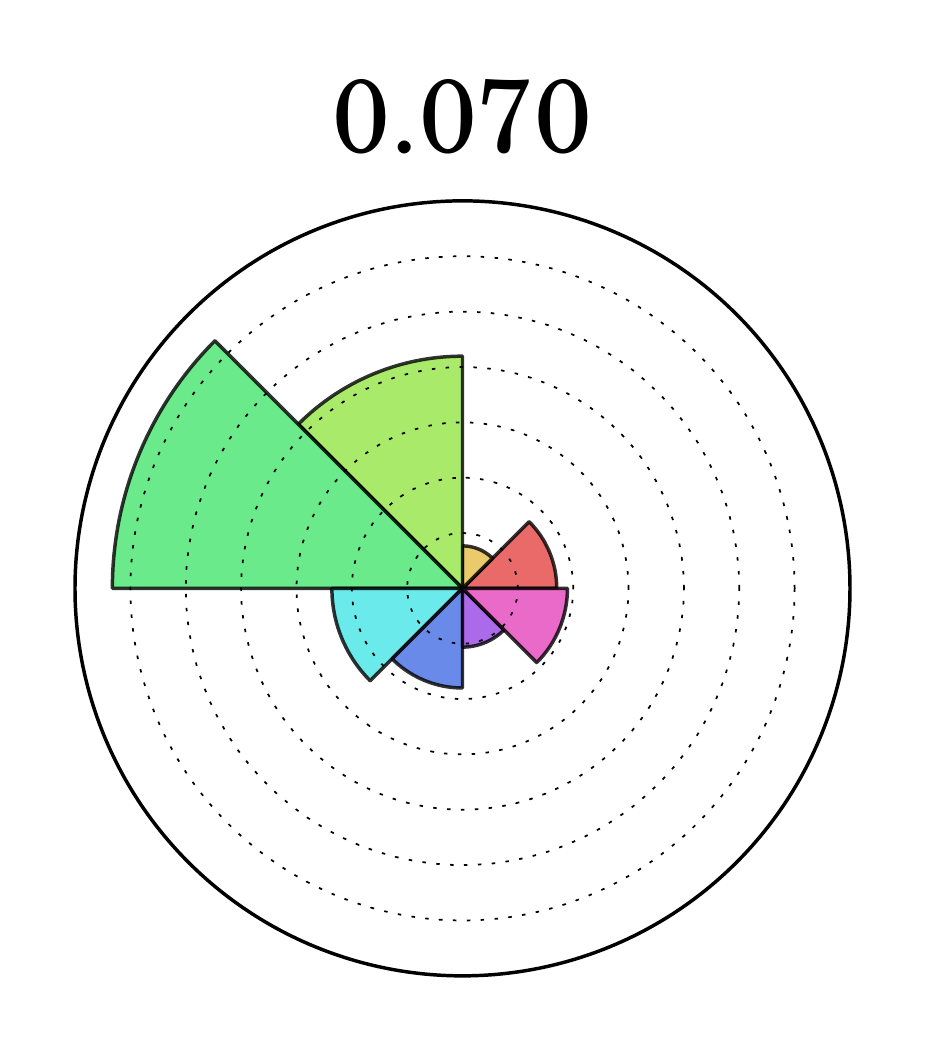}
			\caption*{\hfill Residences \hfill \break}
		\end{subfigure} &
		
		\begin{subfigure}{0.2\columnwidth}
			\includegraphics[width=\columnwidth]{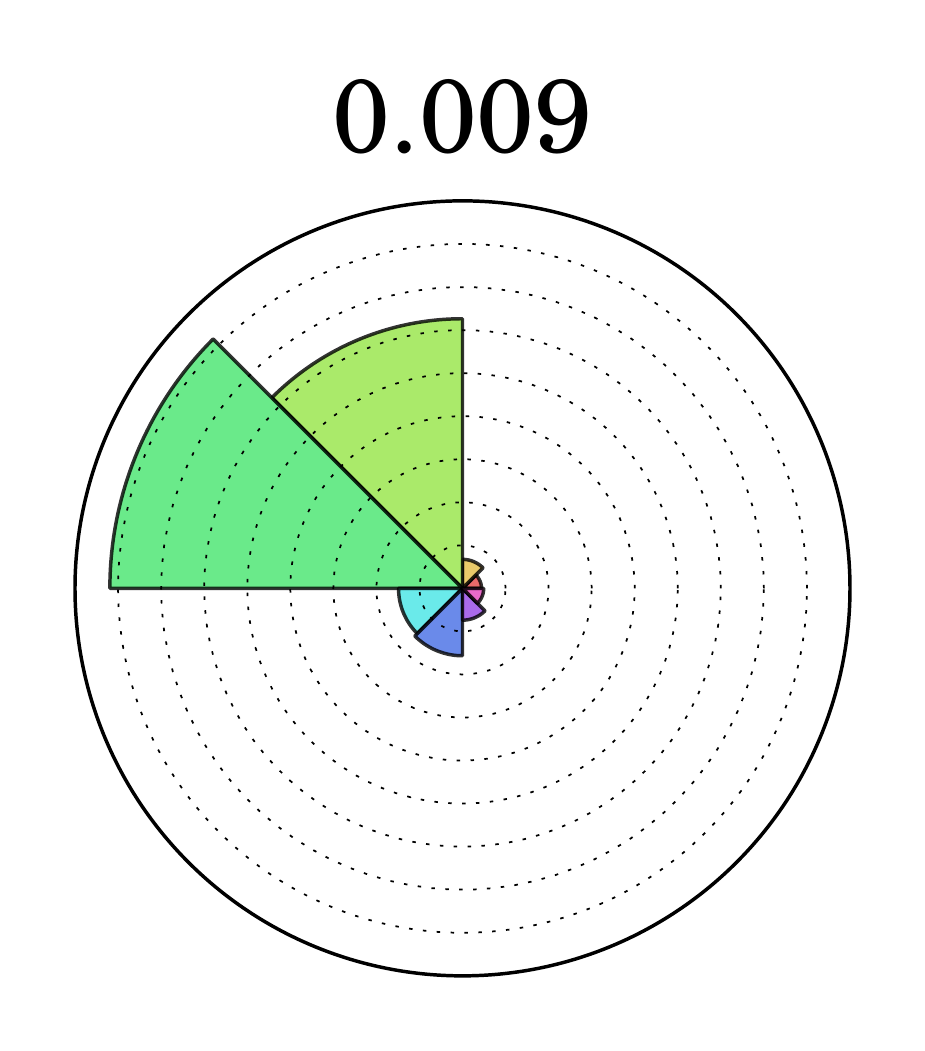}
			\caption*{\hfill Schools \hfill \break}
		\end{subfigure} \\

		\begin{subfigure}{0.2\columnwidth}
			\includegraphics[width=\columnwidth]{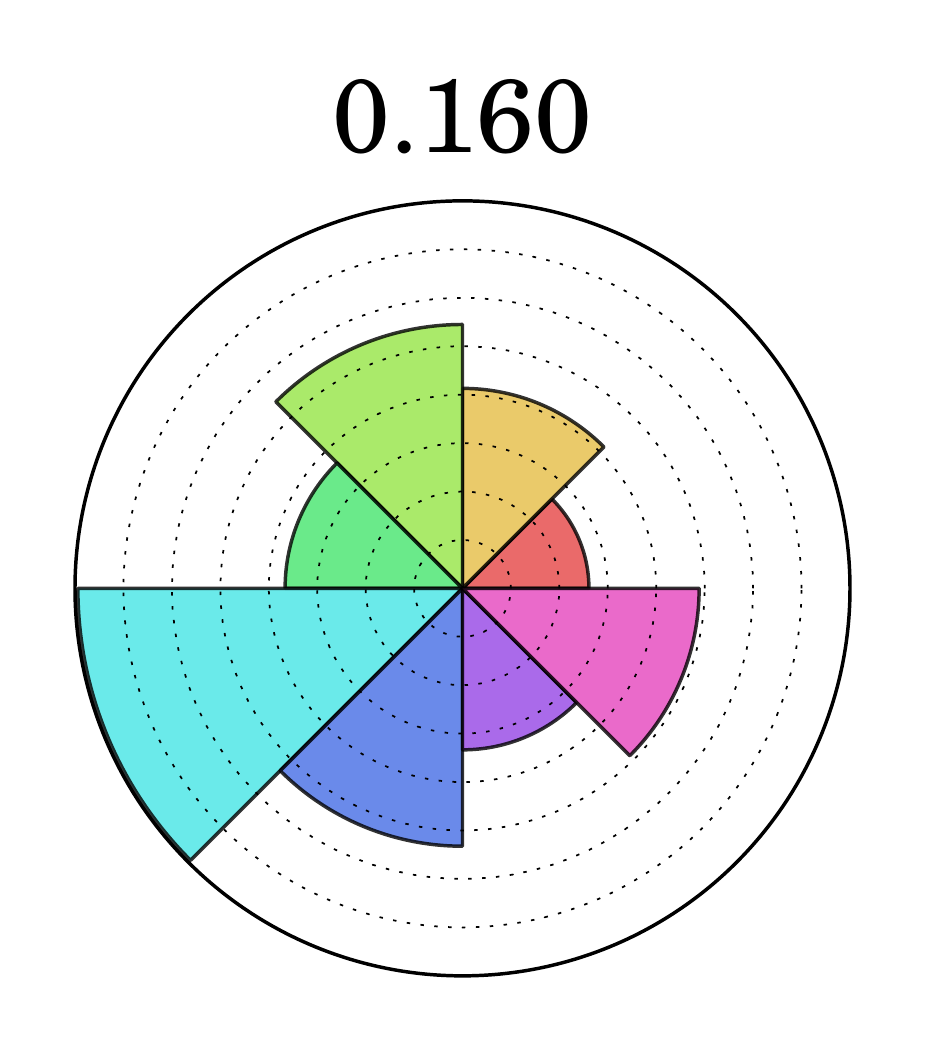}
			\caption*{Shops \& Services}
		\end{subfigure} &
		
		\begin{subfigure}{0.2\columnwidth}
			\includegraphics[width=\columnwidth]{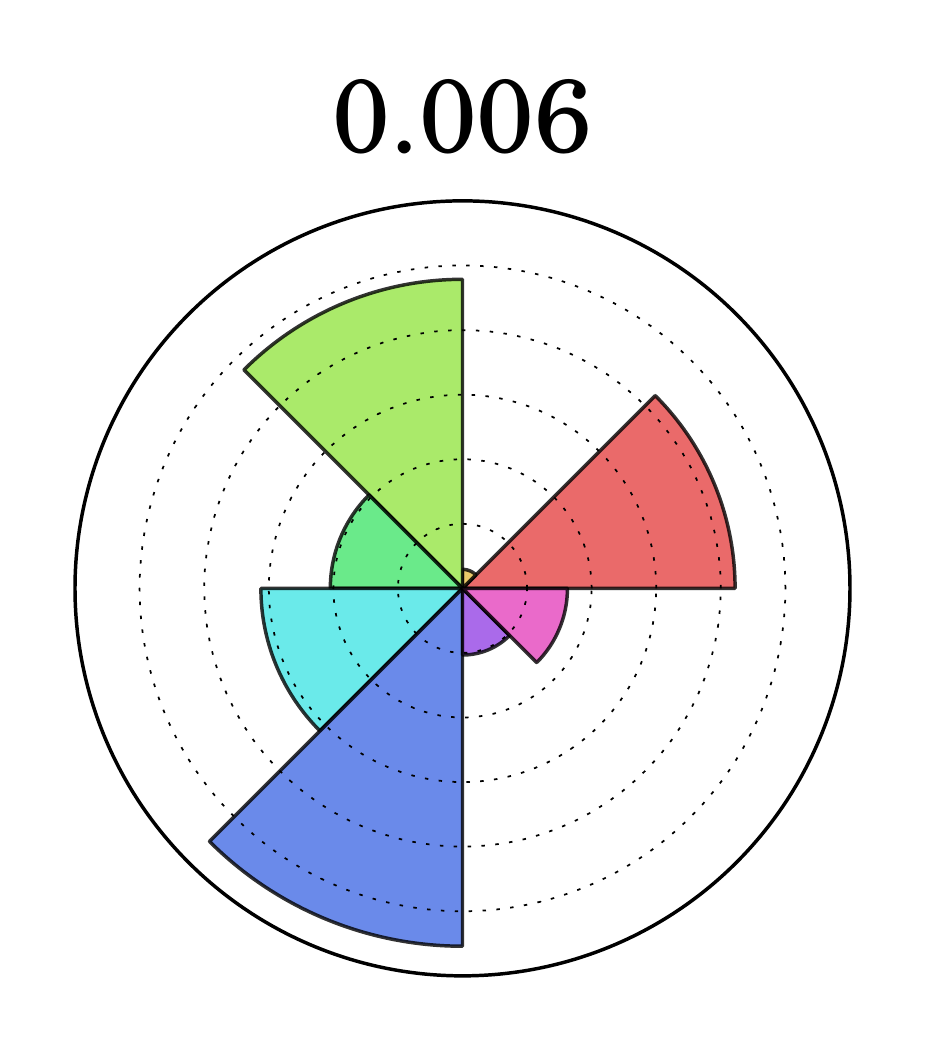}
			\caption*{Spiritual Centers}
		\end{subfigure} &
		
		\begin{subfigure}{0.2\columnwidth}
			\includegraphics[width=\columnwidth]{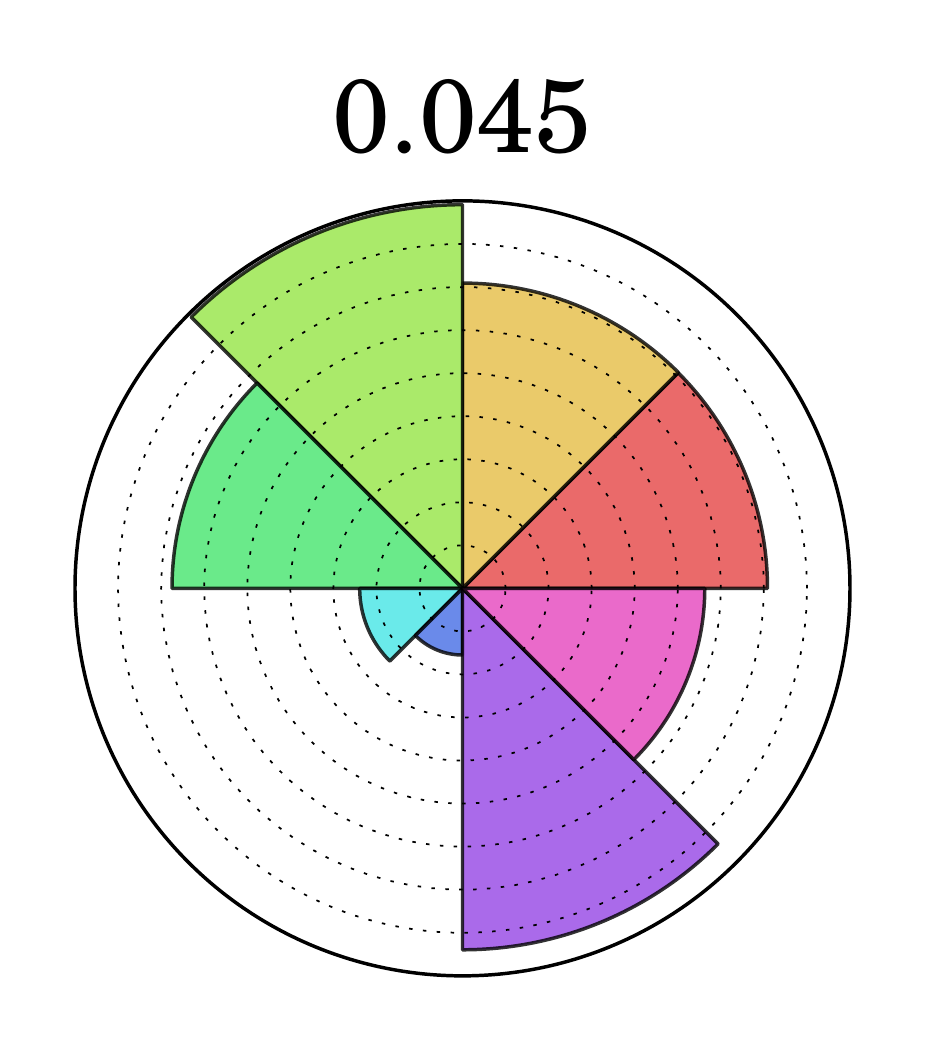}
			\caption*{State Municipalities}
		\end{subfigure} &
		
		\begin{subfigure}{0.2\columnwidth}
			\includegraphics[width=\columnwidth]{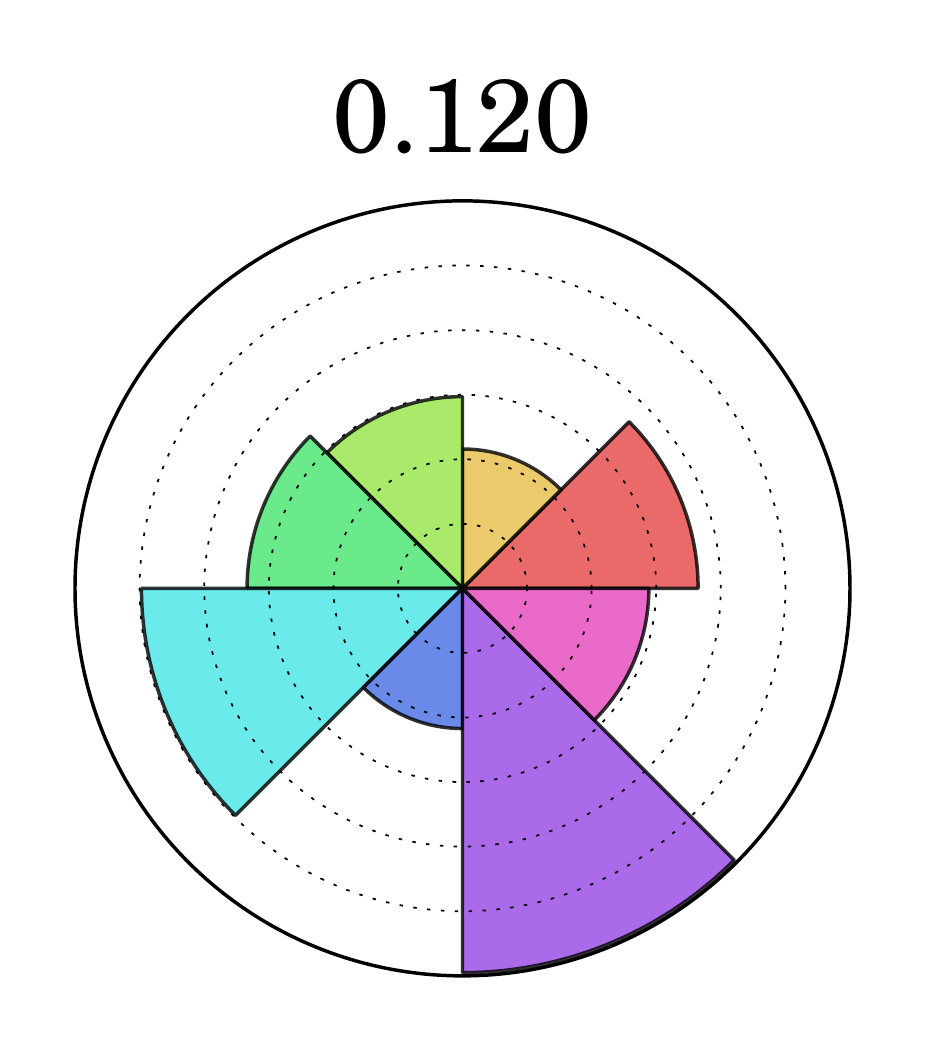}
			\caption*{Travel \& Transport}
		\end{subfigure} \\
	\end{tabular}
	\vspace{0.5em}
	\caption{General category components of each cluster. Each graph shows one component of the vectors shown in Figure~\ref{fig:category_growth_profiles}, averaged over the members of the clusters in Figure~\ref{fig:clustering_map}. The number on the top of each subfigure is the proportion of the vector that the outermost ring represents. e.g. 13\% of all new places in the average country in Cluster 1 are airports.}
	\label{fig:clustering_components}
\end{figure}
\subsubsection*{Results}
Figure~\ref{fig:clustering_map} shows the seven resulting clusters while Figure~\ref{fig:clustering_components} compares the averaged components of each cluster. 
Notably there is a clear intra-cluster spatial correlation, suggesting that geography and, by extension, culture play a significant role in determining the urban activity growth of different cities. 
For example cluster $2$ is predominantly Asian cities, cluster $3$ European cities and cluster $8$ US cities.
As Figure~\ref{fig:clustering_components} 
shows, in Asian cities Spiritual Centers and Food places are common activities in terms of urban development, whereas U.S. Cities focus on Professional and Office spaces, as well as Arts and Entertainment and Sport activities.
European cities on the other hand, appear to invest on Transport and Municipal infrastructure.
While outliers do exist in the composition of the clusters described, those are to some extent expected given that there is some degree of diversity among cities that belong to the same country.
For example, U.S. cities split primarily into two clusters which, nonetheless, feature common characteristics when compared to other clusters, as seen in Figure~\ref{fig:clustering_components}.
Interestingly, many of the Brazilian cities belong to a cluster whose primary feature is activities related to Higher Education reflecting recent policies~\cite{brasil} in the country.
The same cluster also scores highly for development in Athletics and Sports, which we hypothesise is related to the 2014 World Cup, a major sports event hosted in Brazil. 
Finally, Dubai and Queens stand out as Airport hubs due to the presence of Dubai's International Airport (the world's busiest) and LaGuardia Airport that was recently reconstructed.

\subsubsection*{The role of geography}
In order to quantify the spatial correlation we calculated the mean Haversine distance between pairs of cities in the same cluster.
We then randomised the members of each cluster and repeated the calculation. The randomisation process fixed the number of clusters and their sizes. 
Cities were reassigned by first generating a list of all cities from the concatenation of the clusters in a fixed order. 
This list was then randomly permuted, and the new members of each cluster were chosen by repartitioning the list in the same fixed order.
The results in Table~\ref{fig:spatial_correlation_growth_vector} show that the mean intra-cluster pair distance is just over half of the corresponding figure if the cities were randomly assigned to clusters without any consideration to geography.

\begin{table}[ht]
	\centering
	\begin{tabular}{lr}
		\toprule
		Clustering method & Mean distance (km) \\
		\midrule
		Growth vector & 4,868 \\
		Random & 8,752 \\
		\bottomrule
	\end{tabular}
	\caption{Quantifying the spatial correlation for clustering by growth vectors}			\label{fig:spatial_correlation_growth_vector}
\end{table}

\begin{figure}[!ht]
	\centering
	\includegraphics[width=\columnwidth]{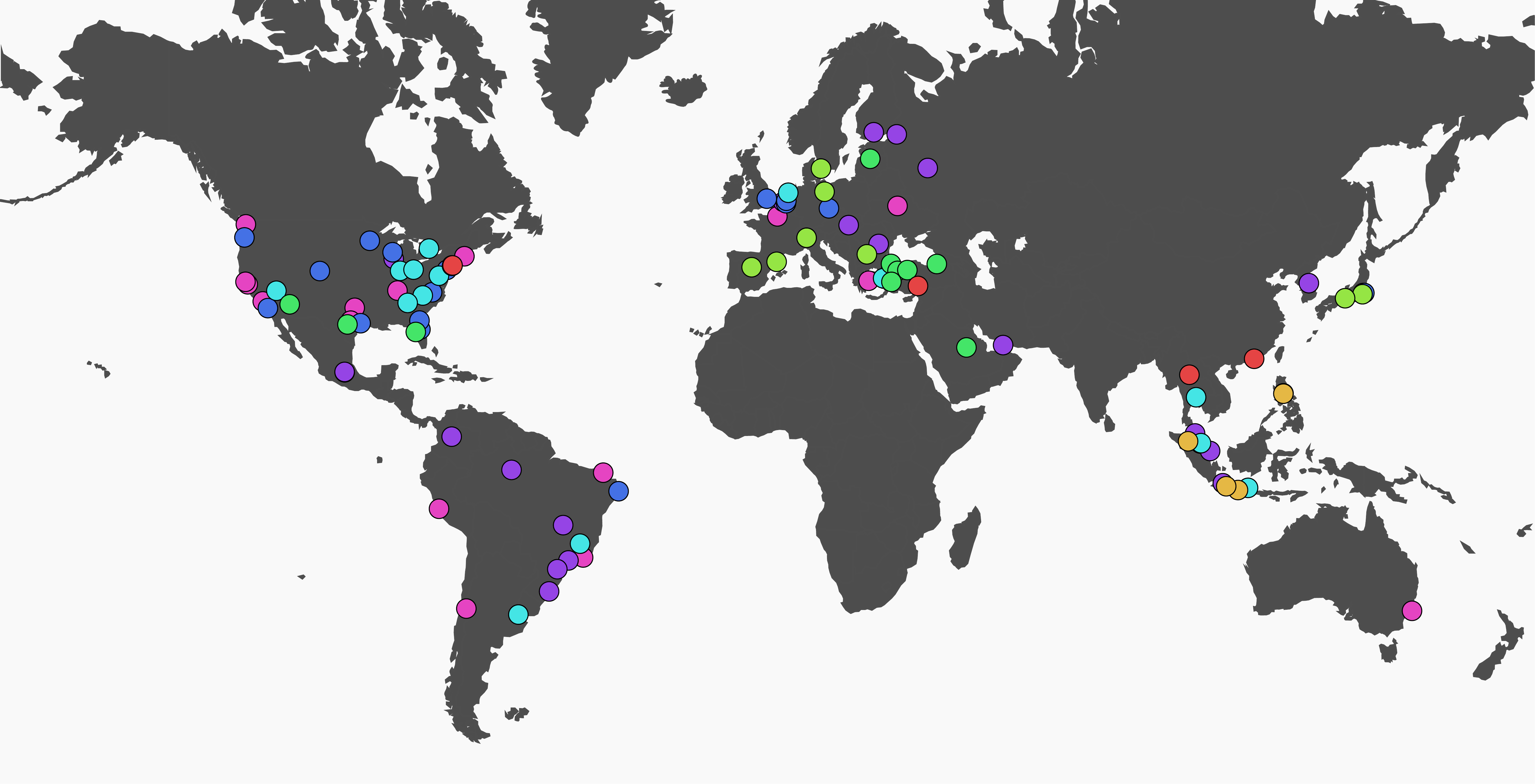}
	
	\vspace{1em}
	
	\begin{tabularx}{\columnwidth}{ccX}
	\toprule
	\# & Colour & Cities \\
	\midrule
	1 & \includegraphics[width=\csize]{cluster1colour2.pdf} & Adana, Borough of Queens, Chiang Mai, Hong Kong \\
	2 & \includegraphics[width=\csize]{cluster2colour2.pdf} & Bandung, Makati City, Medan, Quezon City, Yogyakarta \\
	3 & \includegraphics[width=\csize]{cluster3colour2.pdf} & Barcelona, Berlin, Copenhagen, Madrid, Milano, Osaka, Sofia, Yokohama \\
	4 & \includegraphics[width=\csize]{cluster4colour2.pdf} & Ankara, Denizli, Eski\c{s}ehir, Phoenix, Riga, Riyadh, San Antonio, Tampa, Trabzon, \.{I}stanbul \\
	5 & \includegraphics[width=\csize]{cluster5colour2.pdf} & Amsterdam, Atlanta, Bangkok, Belo Horizonte, Brooklyn, Buenos Aires, Charlotte, Columbus, Indianapolis, Kuala Lumpur, Las Vegas, Petaling Jaya, Surabaya, Toronto, Washington D.C., \.{I}zmir \\
	6 & \includegraphics[width=\csize]{cluster6colour2.pdf} & Antwerpen, Brussels, Chiba, Denver, Houston, Jacksonville, London, Milwaukee, Minneapolis, Orlando, Philadelphia, Portland, Prague, Raleigh, Recife, San Diego, Tokyo \\
	7 & \includegraphics[width=\csize]{cluster7colour2.pdf} & Bogot\'{a}, Bras\'{i}lia, Bucharest, Budapest, Chicago, Curitiba, Dubai, Gent, George Town, Helsinki, Jakarta, Manaus, Mexico City, Moscow, Porto Alegre, Saint Petersburg, Seoul, Singapore, S\~{a}o Paulo \\
	8 & \includegraphics[width=\csize]{cluster8colour2.pdf} & Athens, Austin, Boston, Bursa, Coyoac\'{a}n, Dallas, Fortaleza, Kiev, Lima, Los Angeles, Nashville, New York, Paris, Pineda, Rio de Janeiro, San Francisco, San Jose, Santiago, Seattle, Shah Alam, Sydney \\
	\bottomrule
	\end{tabularx}
	\caption{Relative growth vector clustering results}
	\label{fig:relative_clustering_map}
\end{figure}

Given that the magnitude of the Food component of the city growth vectors dominates other categories, we recalculated the results using \emph{relative} activity growth vectors. 
The relative growth vector represents the comparative growth in the place category rather than the absolute growth. This is defined as:
\begin{equation} \label{eq:relative_growth_vector}
\vec{v}_i = \frac{N_{new}(i)}{N_{old}(i) \sum_j \frac{N_{new}(j)}{N_{old}(j)}}
\end{equation}
For example, although Food previously carried a significant proportion of the original vectors' weight, it only accounts for a small proportion of the relative growth vector as there are a large number of existing Food places.
The results are shown in Figure~\ref{fig:relative_clustering_map}.
Although the number and size of the clusters remains roughly constant, the spatial correlation within the groups largely disappears as Table~\ref{fig:spatial_correlation_relative_growth_vector} shows.
We hypothesise the existing composition of place types in a city is a result of strong cultural and geographical influences.
The relative growth vector, by definition, emphasises more recent trends in urban growth.
It thus captures the short term evolution of urban activities beyond the standard historical and regional patterns.
It is therefore more likely to reflect local organisational factors such as municipal policies or short-duration infrastructure projects.
This type of growth, as indeed Table~\ref{fig:spatial_correlation_relative_growth_vector}  suggests, may be unlikely to have as strong a geographical correlation.

\begin{table}[ht]
	\centering
	\begin{tabular}{lr}
		\toprule
		Clustering method & Mean distance (km) \\
		\midrule
		Relative growth vector & 7,716 \\
		Random & 8,297 \\
		\bottomrule
	\end{tabular}
	\caption{Quantifying the spatial correlation for clustering by relative growth vectors}
\label{fig:spatial_correlation_relative_growth_vector}
\end{table}


\subsection{Detecting surges in urban growth} \label{sec:new_place_spatial_distribution}
So far we have provided a cross city, global overview of urban activity growth patterns.
We have seen that cities which belong to the same country or continent are more likely to feature similar urban activity profiles.
However, the data available through location-based services features very high spatial granularity and the position of real world places nowadays is known with accuracy down to five meters or less~\cite{gpsaccuracy}.
This provides an opportunity to study urban activity growth patterns at an intra-city level.
Motivated by these observations, in this section we ask: \textit{Can data from location-based services can be exploited to identify areas in cities where there is surge in urban development?} 

\begin{figure}[h!]
    \centering
    
    \newcommand*{\figwidth}{0.49\columnwidth}
    
    \begin{subfigure}{\columnwidth}
    	\includegraphics[width=\figwidth]{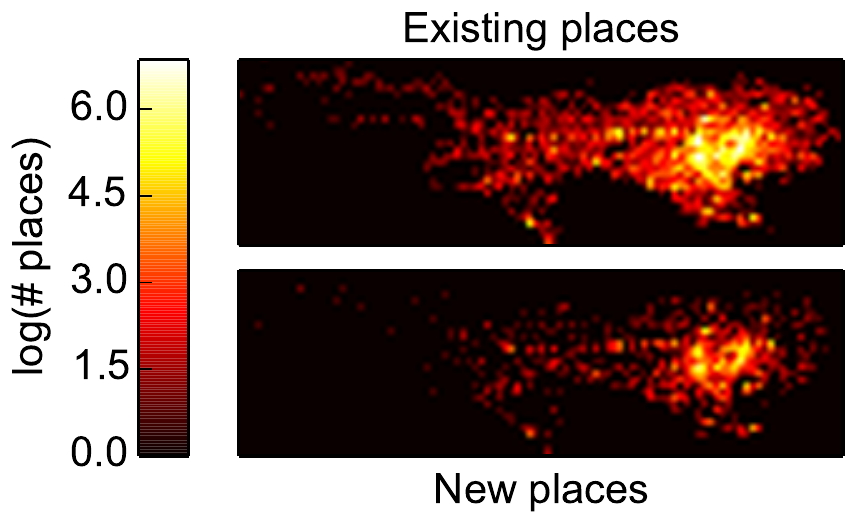}
    	\hfill
    	\includegraphics[width=\figwidth]{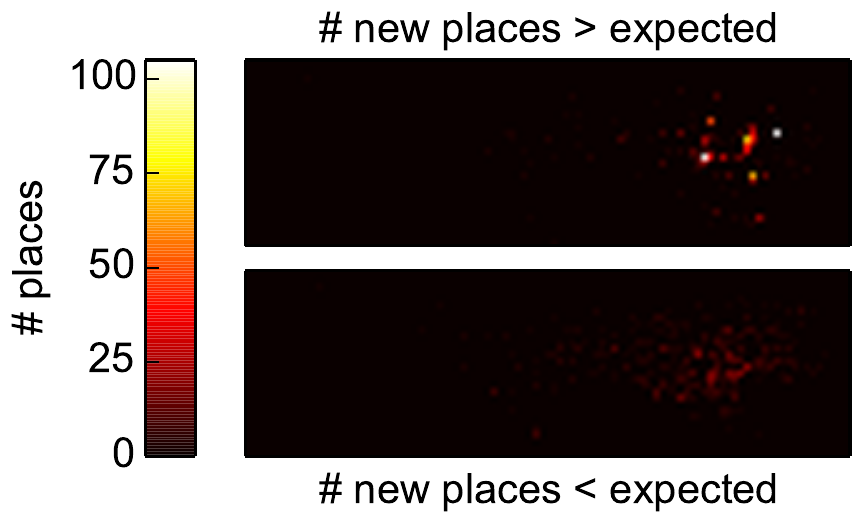}
    	\caption{Tokyo}
    \end{subfigure}
	
	\vspace{1em}
	
    \begin{subfigure}{\columnwidth}
    	\includegraphics[width=\figwidth]{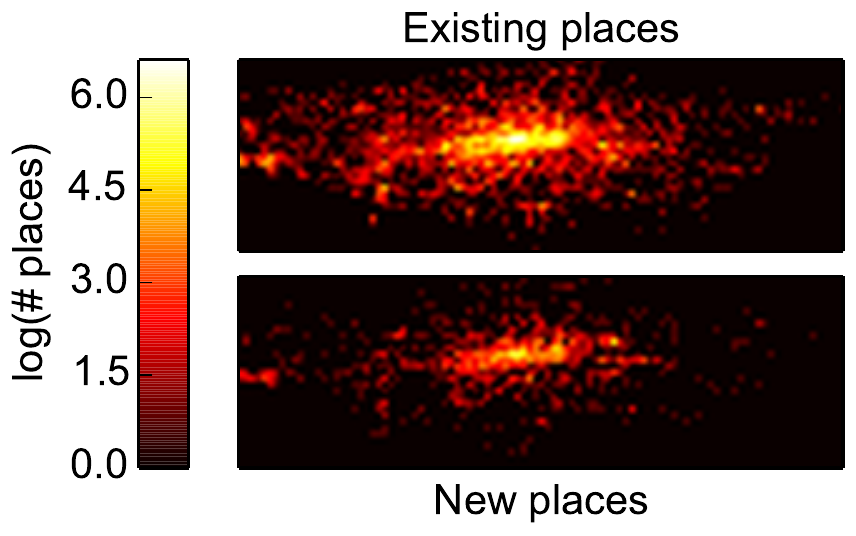}
    	\hfill
    	\includegraphics[width=\figwidth]{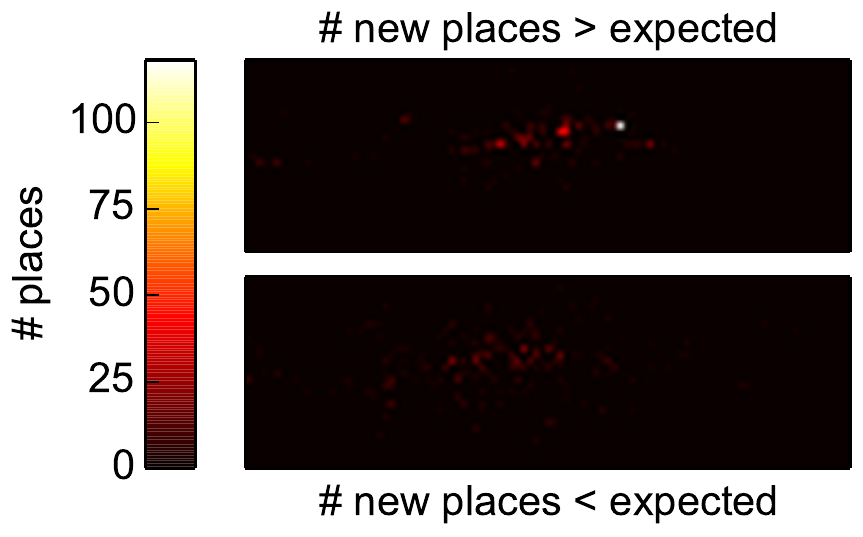}
    	\caption{London}
    \end{subfigure}
    
    \begin{subfigure}{\columnwidth}
    	\includegraphics[width=\figwidth]{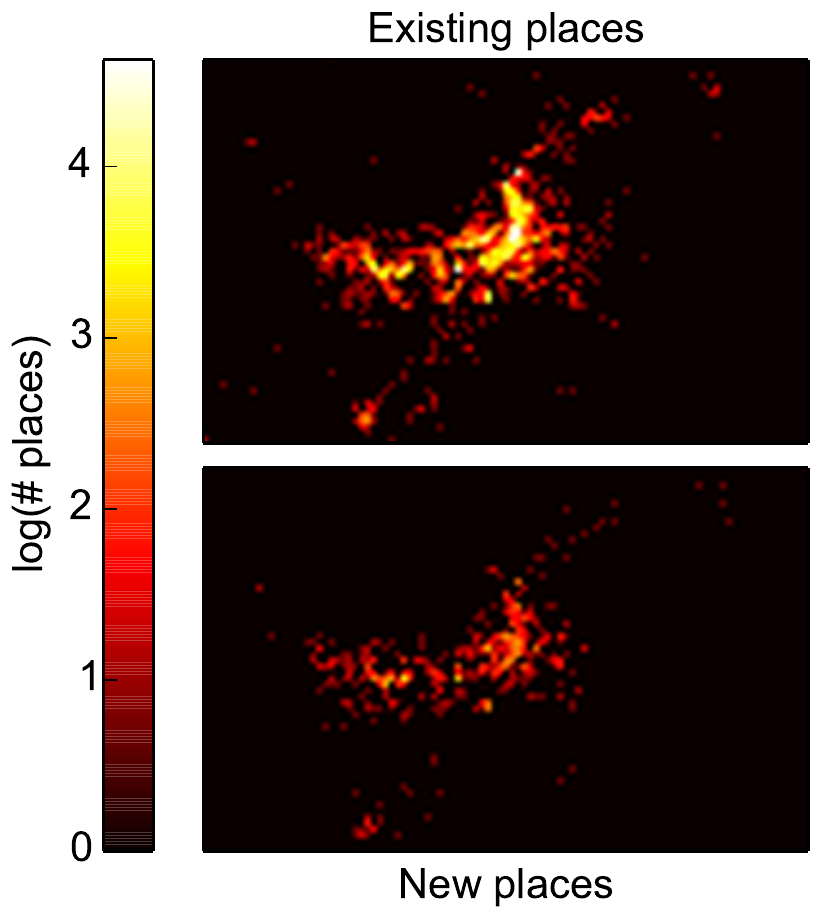}
    	\hfill
    	\includegraphics[width=\figwidth]{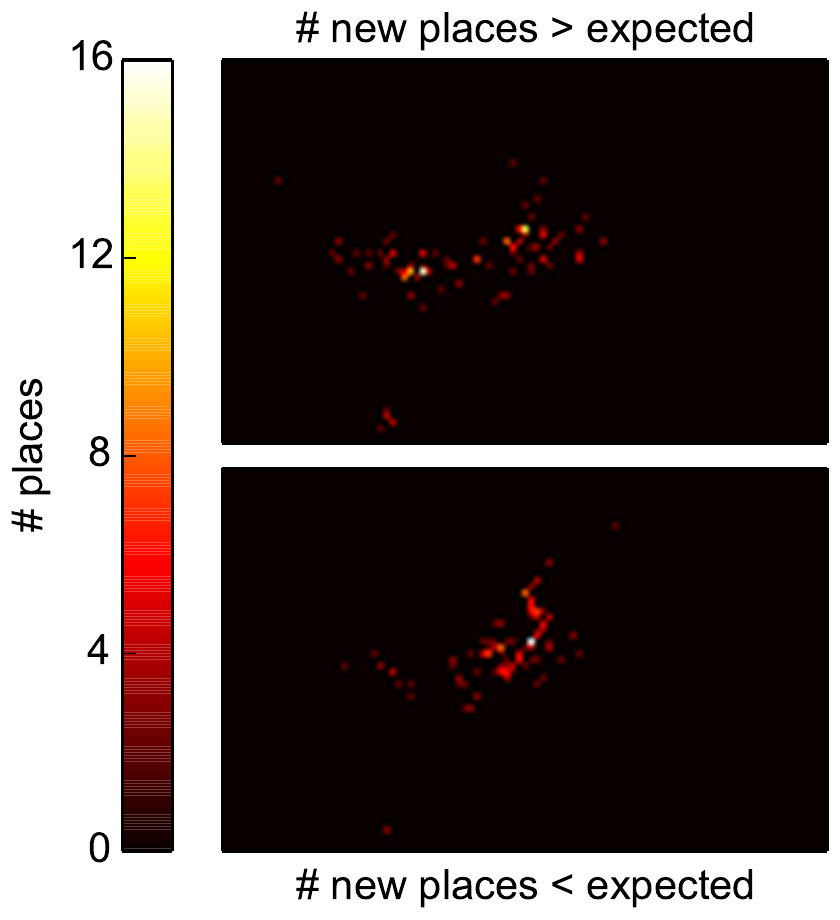}
    	\caption{Brasilia}
    \end{subfigure}
	
	\vspace{1em}
    
    \begin{subfigure}{\columnwidth}
    	\includegraphics[width=\figwidth]{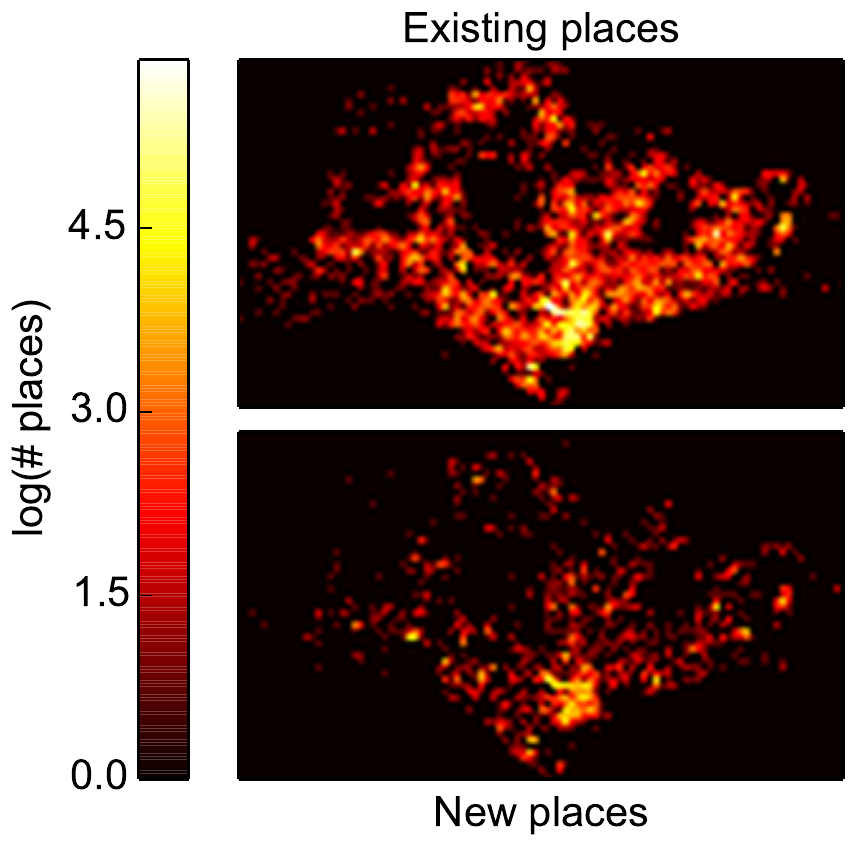}
    	\hfill
    	\includegraphics[width=\figwidth]{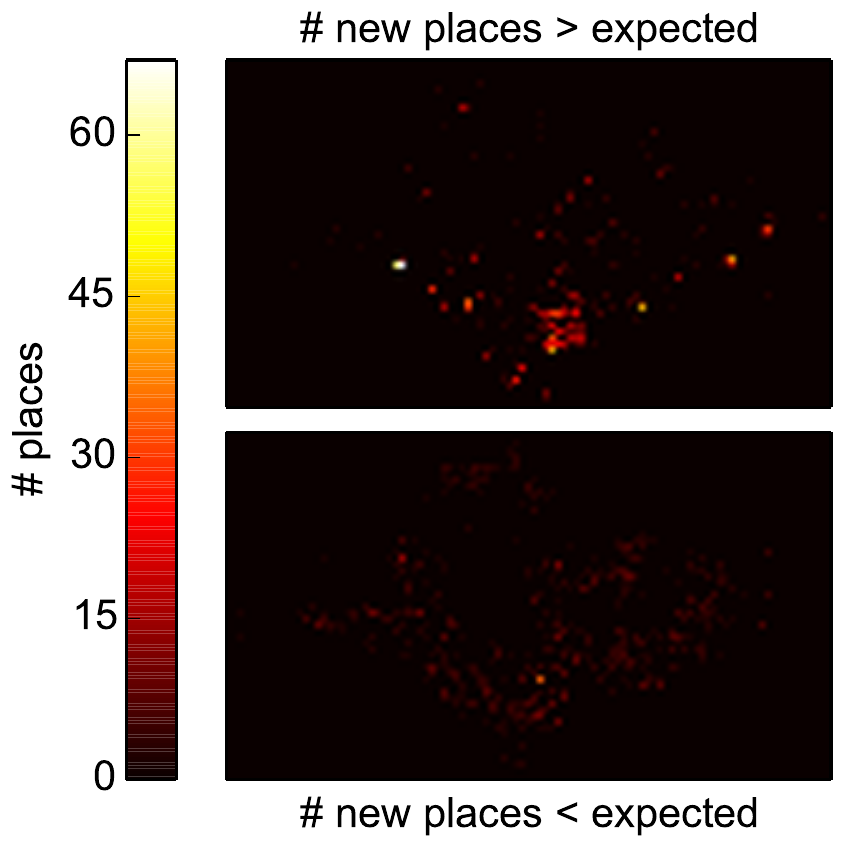}
    	\caption{Singapore}
    \end{subfigure}
    \caption{Intensity plots of new places in cities. Upper left: log(existing places). Lower left: log(new places). Upper right: number of new places less than expected. Lower right: number of new places greater than expected.}
    \label{fig:city_intensity_plots}
\end{figure}

To answer this, we first investigate the spatial distribution of new places within cities. Initially we divide each city up into a regular 100x100 grid and measure the urban activity on a per cell basis.
We create spatial intensity plots for the distribution of both existing venues and new venues. Examples of these can be seen on the left-hand side of Figure~\ref{fig:city_intensity_plots}. 
Having observed that the spatial distribution of new venues does not always coincide with that of existing venues, we now present a methodology for comparing the difference between them.

Our null model randomly redistributes new venues so that the density of new venues in each cell is proportional to the density of the cell's existing venues:
\begin{equation} \label{eq:null_distribution_model}
n_{i,j}^\text{null} = \frac{n_{i,j}^\text{existing} n^\text{new}}{n^\text{existing}}
\end{equation}
where $n_{i,j}^{null}$ represents the number of new venues in cell $(i,j)$, and $n^\text{new}$ and $n^\text{existing}$ are the total number of new and existing venues in the city respectively.
By subtracting the true distribution of new venues from that of the null model (Equation~\ref{eq:spatial_model}), we highlight where new place growth is higher and lower than expected. 
\begin{equation} \label{eq:spatial_model}
v_{i,j} = n_{i,j}^\text{null} - n_{i,j}^\text{new}
\end{equation}

Figure~\ref{fig:city_intensity_plots} shows the results for four example cities. The most obvious trend is that areas of higher than expected levels of new places are highly concentrated, as opposed to areas of lower than expected development which are scattered broadly. This pattern appears to reflect well the fact that development happens strategically as dictated by the city's authorities, whereas deprivation in development is not planned. Interestingly, from these plots it is also possible to pick out several notable developments and events. In London, we our methodology is able to detect the development of the The Olympic Village built for the 2012 Olympic games in Stratford (white dot). Another example is Brasilia one can spot The Man\'{e} Garrincha National Stadium rebuilt for the 2014 Football World Cup (rightmost yellow dot). There examples demonstrate how the digital datasets emerging from location-based technologies can become an important tool for monitoring urban development on a large scale and with unprecedented geographic accuracy.


\section{Micro-scale analysis} \label{sec:impact}
We have looked at how the emergence of new urban activities in cities can be tracked and showed how the urban development profiles result in clusters of cities that are geographically and culturally proximate.
Further we have demonstrated a method that allows for the detection of surge in urban development across a city's territory. 
In other words, we have explored the effects of new place creation in cities as a whole.

We now turn our investigation to how a new venue can influence traffic between other venues in the local area.
Answering this question is of interest for the sake of understanding the impact of investment on certain urban activities on a large scale.
It has also implications for important problems in retail geography~\cite{karamshuk13, coelho1976optimum, jensen06}, such as deciding where to open a new business in the city. 

A critical question to ask in this setting is whether a new venue cooperates with nearby venues by increasing footfall in the area, or instead competes, with customers being redirected from existing venues to the new venue's premises.
While various theories have been proposed on the subject since the 1920s~\cite{hotelling1929stability} the new generation of data available from location-based services offers the opportunity to empirically investigate these phenomena on a large scale.
In particular we ask \textit{How does the opening of different venue categories affect foot traffic to nearby venues?}



Our core assumption is that the real popularity of a venue, as measured by the number of visitors, is correlated with the number of check-ins by mobile users.
Of course, a venue's popularity is a complicated function of many variables including, but not limited to: consumer awareness, one-off events and even the state of the country's economy.
It is therefore important to note that when measuring the impact on a given pair of venues, the new one and the existing one, we do not imply a causal effect. 
However aggregating impact values over a sufficiently large number of pairs and using millions of records of mobility data, should allow us to discern overall trends and obtain informative signals on the interaction and reciprocal influence of urban activities in neighbourhoods.

\subsubsection*{A definition of ``impact"}
To do so we define an appropriate \emph{impact} measure. 
A new venue's impact on an existing venue is defined to be the ratio between the normalised average number of transitions per month involving the existing venue in the $6$ month time period before and after the new venue is added to the database.
Formally the impact metric is defined as:
\begin{equation} \label{eq:definition_of_impact}
\text{impact}(e, m) = \frac{\sum_{d=m}^{m+5} \frac{t_{e}(d,d+1)}{n_{e}(d,d+1)}} {\sum_{d=m}^{m-5} \frac{t_{e}(d-1,d)}{n_{e}(d-1,d)}}
\end{equation}
where $e$ is the existing venue, $m$ is the month when the new venue was added to the database, $t_{e}(x,y)$ is the number of transitions involving $e$ that took place between dates $x$ and $y$ and $n_{e}(x,y)$ is a normalisation constant for venue $v$ between dates $x$ and $y$.
The normalisation constant $n_v(x,y)$ takes into account the overall change in monthly Foursquare usage on a city-by-city basis.
It also normalises with respect to the monthly transition counts for all venues in the city of the same specific category.
This should remove both seasonal factors (e.g. Ice Cream Parlours) and categories undergoing dramatic changes in popularity.

For example an impact value of 1.28 represents a 28\% increase in the average normalised number of transitions per month after the new venue opens, while 0.92 represents an 8\% decrease.
We also remove from our analysis any new-old venue pairs that have less than 6 months of transition counts, either before or after the opening of the venue.
Finally, the 6 month average was chosen as a compromise value.
Given that the dataset covers a time period of four years, too long a measurement period would remove a sizeable portion of the dataset from consideration, while too short a period would result in significant noise being added to the impact values.

\subsection{Homogeneous urban activity interactions} \label{sec:homogeneous_pairs}

We now explore the effects of new places opening up in close proximity to existing places that are of the same category.
That is, they represent the same
urban activity.
As an initial study it has the advantage that one would expect minimal differences between the times venues of the same type are active. 
Naturally, if venues don't share similar opening times, they are unlikely to interact directly.
A nightclub and a primary school, even if they are on the same street, are unlikely to exchange customer flows or disrupt each other.
As a result, this approach removes an important variable from consideration in the analysis.
In this analysis we take advantage of the depth of detail present in the Foursquare dataset and consider more refined place types.
Whereas in the previous analysis we viewed, for example, Italian Restaurants and Japanese Restaurants as part of the Food category~\footnote{\url{https://developer.foursquare.com/categorytree}}, we now consider them as distinct categories in their own right.

\subsubsection*{Case studies: Burger Joints, Bookstores and Airport Gates}

\begin{figure}[h!]
    \begin{subfigure}{\columnwidth}
    	\centering
    	\includegraphics[width=0.85\columnwidth]{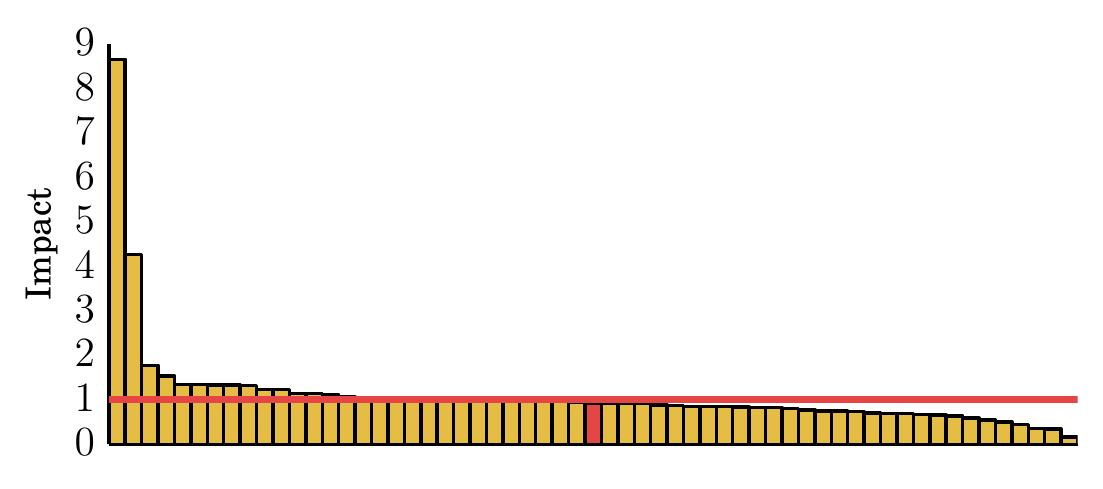}
    	\caption{Burger joints (59 pairs)}
    \end{subfigure}

	\vspace{1em}    
    
    \begin{subfigure}{\columnwidth}
    	\centering
    	\includegraphics[width=0.85\columnwidth]{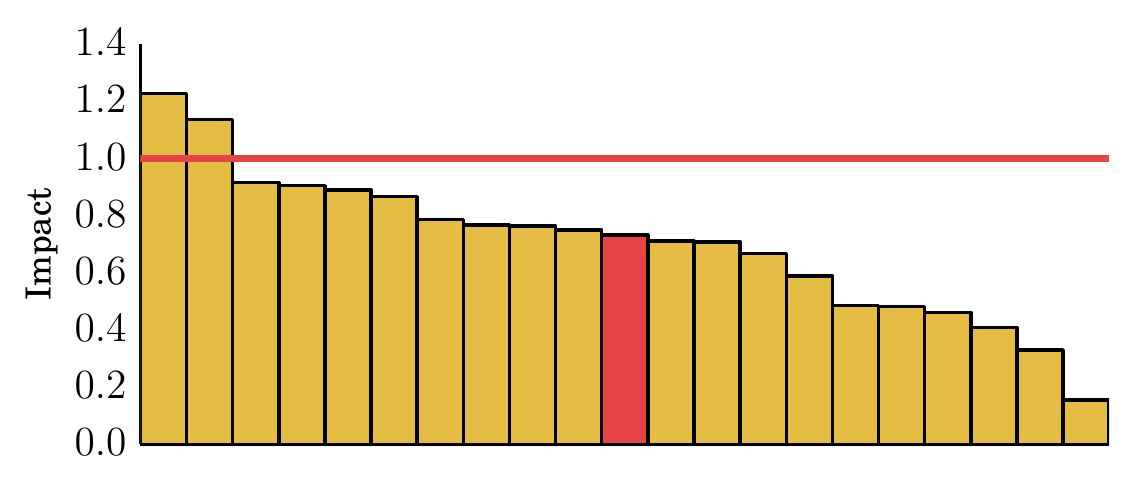}
    	\caption{Bookstores (21 pairs)}
    \end{subfigure}
    
    \vspace{1em}
    
    \begin{subfigure}{\columnwidth}
    	\centering
    	\includegraphics[width=0.85\columnwidth]{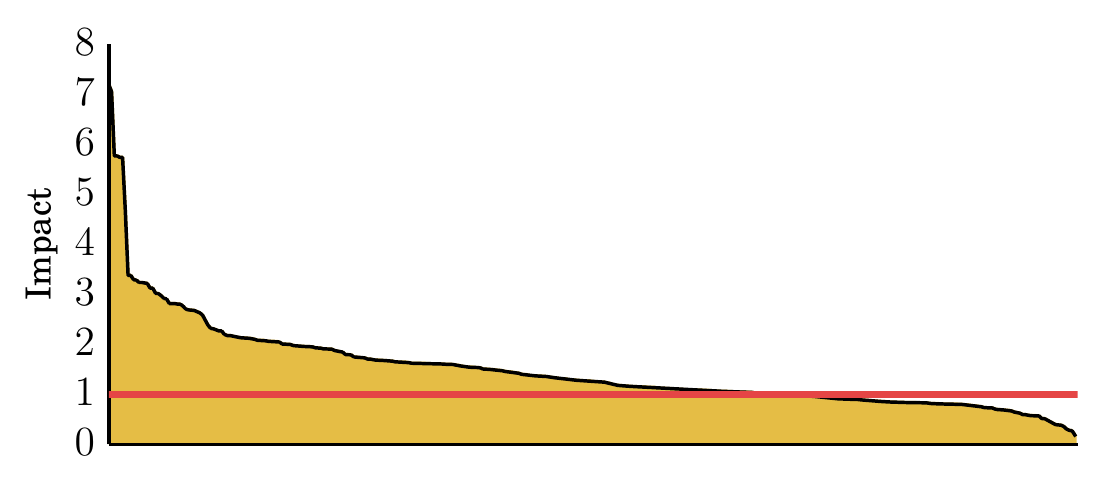}
    	\caption{Airport gates (352 pairs)}
    \end{subfigure}
    \caption{Impact values of all new-existing pairs in London. Each bar represents a new venue opening up within 500m of an existing venue. An impact value of greater than 1.0 indicates a positive effect on the existing place type, less than 1.0 a negative impact. The red bar indicates the median pair.}
    \label{fig:case_study_impacts}
\end{figure}

As an initial case study we investigate the categories ``Burger joints", ``Bookstores" and ``Airport Gates" in London. We choose this particular subset of categories as Burger Joints are the fastest growing specific category in London venues, Bookstores are a category we hypothesise to be naturally competitive and Airport Gates should be naturally cooperative (as new gates are indicative of expansion in the airport capacity).
Using KD-trees, an efficient spatial look-up algorithm \cite{bentley75}, we identified all instances of a new venue opening up within 500m of an existing venue of the same specific category. For each pair we then calculated the impact of the new venue on the existing venue as defined in Equation~\ref{eq:definition_of_impact}.

The results for the three categories are shown in Figure~\ref{fig:case_study_impacts}. The mean change in average monthly transitions for an existing burger joint after a new burger joint opened up within $500$m is an increase of $8.5$\%.
At first glance this suggests that burger joints may have a positive effect on footfall to the area.
However this mean increase is the result of two outliers. 
The median change is in fact a decrease of $12.4$\%.
This suggests that new Burger joints in fact steal customers from existing Burger joints. 
As hypothesised, $19$ out of the $21$ existing book stores experienced a significant decrease in traffic after another book store opened up nearby. Similarly the majority of Airport gates experience an increase in traffic after a new Airport gate opens up nearby. 

\subsubsection*{Top competitive and cooperative categories} \label{sec:top_homogeneous_comp_coop_types}

Given the encouraging results from the case studies, we performed similar analysis for all categories in the dataset that had at least $10$ new-old pairs. Given the high variance observed in Figure~\ref{fig:case_study_impacts}, in the following analysis we rank the categories by both the median and the mean impact. This is because the mean can often be misleading as a measure of typical behaviour in distributions with high variance.

\begin{table}[ht]
	\begin{minipage}{.485\linewidth}
      	\centering
        \begin{tabular}{llc}
			\toprule
			\# & Category & Median \\
			\midrule
			1. & Coworking Spaces & 0.415 \\
			2. & Tapas Restaurants & 0.632 \\
			3. & Grocery Stores & 0.636 \\
			4. & Cosmetics Shops & 0.666 \\
			5. & Train stations & 0.673 \\
			6. & Pharmacies & 0.673 \\
			7. & Salad Shop & 0.681 \\
			8. & Sandwich Places & 0.720 \\
			9. & Salons/Barbershops & 0.724 \\
			10. & Offices & 0.725 \\
			\bottomrule
		\end{tabular}
		\caption{Top competitive categories by median impact}
		\label{tab:topMedianHomoComp}
    \end{minipage}%
    \hfill
    \begin{minipage}{.485\linewidth}
      	\centering
      	\begin{tabular}{llc}
			\toprule
			\# & Category & Mean \\
			\midrule
			1. & Grocery Stores & 0.678 \\
			2. & Bookstores & 0.702 \\
			3. & Pharmacies & 0.727 \\
			4. & Seafood Restaurants & 0.753 \\
			5. & Asian Restaurants & 0.761 \\
			6. & Ice Cream Shops & 0.813 \\
			7. & Capitol Buildings & 0.825 \\
			8. & Salons/Barbershops & 0.834 \\
			9. & Tapas Restaurants & 0.838 \\
			10. & Gastropubs & 0.842 \\
			\bottomrule
		\end{tabular}
		\caption{Top competitive categories by mean impact}
		\label{tab:topMeanHomoComp}
    \end{minipage}
\end{table}
Tables~\ref{tab:topMeanHomoComp} \&
\ref{tab:topMedianHomoComp} show the top $10$ categories that have the lowest median and mean impact. These can be characterised as the categories that are the most competitive. Food related categories make up a significant percentage of the most competitive places, with 
Pharmacies and Bookstores also being present.

\begin{table}[ht]
    \begin{minipage}{.49\linewidth}
      	\centering
        \begin{tabular}{cll}
			\toprule
			\# & Category & Median \\
			\midrule
			1. & Neighbourhoods & 1.480 \\
			2. & Turkish Restaurants & 1.354 \\
			3. & Gardens & 1.341 \\
			4. & Monuments & 1.337 \\
			5. & Plazas & 1.285 \\
			6. & Tea Rooms & 1.264 \\
			7. & Airport Gates & 1.260 \\
			8. & Churches & 1.182 \\
			9. & Museums & 1.115 \\
			10. & Bus Stations & 1.106 \\
			\bottomrule
		\end{tabular}
		\caption{Top cooperative categories by median impact}
		\label{tab:topMedianHomoCoop}
    \end{minipage} 
    \hfill
	\begin{minipage}{.49\linewidth}
      	\centering
      	\begin{tabular}{cll}
			\toprule
			\# & Category & Mean \\
			\midrule
			1. & Museums & 3.236 \\
			2. & Roads & 1.902 \\
			3. & Convention Centers & 1.890 \\
			4. & Plazas & 1.736 \\
			5. & Academic Buildings & 1.543 \\
			6. & Airport Gates & 1.492 \\
			7. & Neighbourhoods & 1.489 \\
			8. & Bus Stations & 1.467 \\
			9. & Tea Rooms & 1.380 \\
			10. & Turkish Restaurants & 1.365 \\
			\bottomrule
		\end{tabular}
		\caption{Top cooperative places by mean impact}
		\label{tab:topMeanHomoCoop}
    \end{minipage}%
\end{table}

Tables~\ref{tab:topMeanHomoCoop} \& \ref{tab:topMedianHomoCoop} show the top $10$ categories that have the highest median and mean impact. 
These can be characterised as the categories that are the most cooperative. 
Many of the entries are to be expected. 
When a new museum opens next to an existing museum it increases the likelihood the desirability of the area. 
Additionally many physically close museums run promotions with joint tickets for both venues. 
Likewise a new academic building opening is likely to be part of the same educational establishment of nearby academic buildings and hence likely foreshadows an expansion in the capacity of that establishment.
We also note that there is a high degree of agreement between the median and the mean, with 7 out of the 10 entries being shared between both lists. 

The oddity in the list of cooperative places is the Turkish Restaurant category. 
Restaurants (and food categories in general) tend to be competitive as Table \ref{fig:restaurant_impacts} shows. 
After all a customer is unlikely to eat twice, no matter how many restaurants are in an area. 
We explore this anomaly further in the next section.
 
\begin{table}[ht]
	\centering
	\begin{tabular}{lcc}
		\toprule
		Restaurant type		& Median impact & Mean impact\\
		\midrule
		Tapas 				& 0.632 		& 0.838\\
		Asian 				& 0.731 		& 0.761\\
		Seafood 			& 0.735 		& 0.753\\
		Mediterranean 		& 0.807 		& 1.298\\
		Chinese 			& 0.811 		& 0.868\\
		Indian 				& 0.827 		& 0.868\\
		Vietnamese 			& 0.844 		& 0.901\\
		Japanese 			& 0.874 		& 0.886\\
		Sushi 				& 0.886 		& 0.939\\
		Mexican 			& 0.908 		& 0.895\\
		Vegetarian / Vegan 	& 0.925 		& 0.874\\
		Korean 				& 0.929 		& 0.998\\
		Thai 				& 0.945 		& 1.023\\
		French 				& 0.974 		& 0.957\\
		American 			& 0.997 		& 1.012\\
		Italian 			& 1.007 		& 1.065\\
		Middle Eastern 		& 1.115 		& 1.233\\
		\textbf{Turkish}	& \textbf{1.354}& \textbf{1.365} \\ 
		\bottomrule
	\end{tabular}
	\caption{Restaurant categories ranked by median impact.}
	\label{fig:restaurant_impacts}
\end{table}

\subsubsection*{Explaining Cooperation Between Turkish Restaurants} \label{sec:turkish_jensen_correlation}
In the previous section we showed that Turkish Restaurants showed a surprising degree of cooperation, with existing Turkish restaurants experiencing a median increase in customers of $35.4$\% in the six months after a new Turkish restaurant opened nearby.
A potential social hypothesis for this is that the Turkish community forms highly concentrated enclaves in London. Under this hypothesis, more ethnic Turkish shops and restaurants open up in existing Turkish communities, creating an ecosystem effect, and further increasingly the likelihood of new Turks settling in the area.

If this hypothesis holds then you would expect very tight clustering of Turkish Restaurants in limited geographical areas.
To test this we turn to Jensen's quality metric \cite{jensen06}, which measures whether the local density of the type of venues is greater or lesser than expected if the spatial distribution of place types was a product of a random process. Equation~\ref{eq:intra_coeff} defines his intra-coefficients for places of the same type.
\begin{equation} \label{eq:intra_coeff}
M_{A} = \frac{N_t - 1}{N_A(N_A-1)} \sum_{i=1}^{N_A} \frac{N_A(A_i,r)}{N_t(A_i,r)}
\end{equation}
where $N_t$ is the total number of places, $N_A$ is the number of places of type $A$, $N_t(A_i,r)$ is the total number of places within a radius $r$ of the $i$th place of type $A$ and $N_X(A_i,r)$ is the number of places of type $X$ within a radius $r$ of the $i$th place of type $A$.
\begin{figure}[ht]
    \centering
    \includegraphics[width=0.85\columnwidth, draft=false]{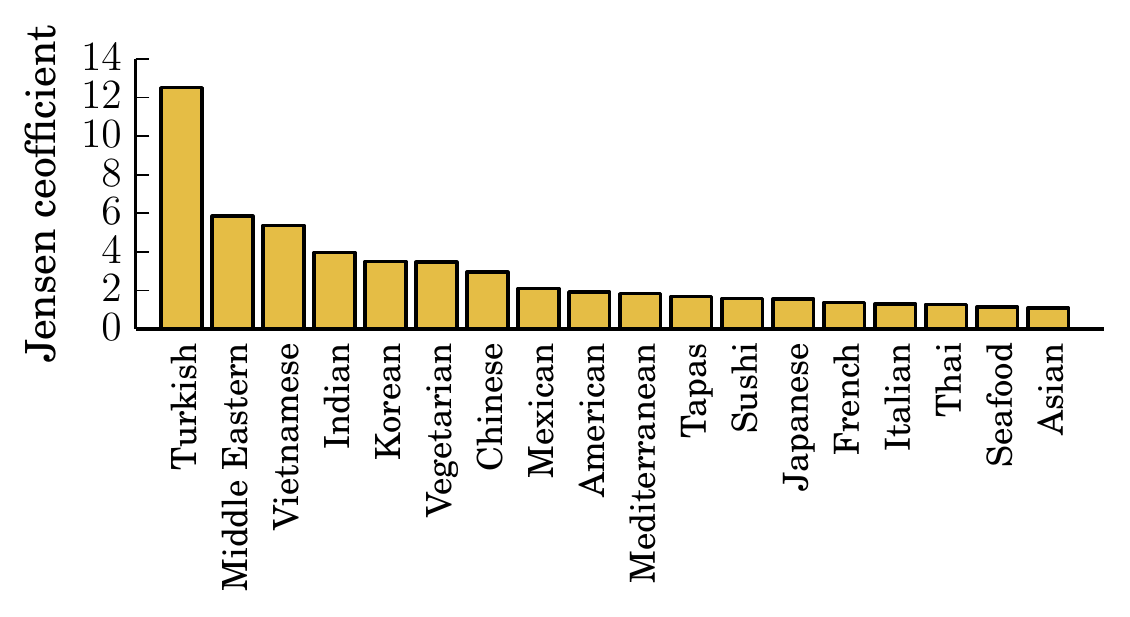}
    \caption{Jensen quality for various restaurant types in London}
    \label{fig:restaurants_jensen_quality}
\end{figure}
Figure 7 plots Jensen's intra-coefficients for each specific Restaurant sub-category using a radius of $500$m. Turkish restaurants are on average twelve times more densely clustered than would be expected if they were distributed at random.
This is far higher then the other restaurant types, confirming our hypothesis that an enclave effect in the Turkish community is affecting the placement of new Turkish Restaurants.

\section{Discussion and Concluding Remarks}
\label{sec:discussion}
In our work we have exploited data sourced from location-based services to reveal urban growth patterns.
Tracking these patterns on a global scale, as our cross-city clustering analysis has shown, can reveal interesting trends in urban development where the role of geography and culture remains important even in an era where centralised city planning is dominant.
However, as our intra-city analysis has shown, confirming recent proposals~\cite{batty2013big}, the power of big location data could allow the detection of large on-going projects across the urban territory. 
This signifies the potential of developing modelling frameworks and software tools that could allow the monitoring of such projects and thus enable more effective city governance. 
The study of the interactions of urban activities as proxied by Foursquare place categories in terms of foot traffic, not only provides more evidence along this direction, but gives rise to the opportunity for the study of the impact of urban growth on nearby places and, in particular, retail establishments.
Previous research has suggested~\cite{karamshuk13, jensen06} that the growth and success of the latter category, appears to be closely related to the formation of the neighboring urban ecosystem.

Overall, our findings complement the current research activities in urban data science.
New layers of data are brought together to understand better urban processes and city life.  
From user surveyed data that help us augment our perspective on urban space~\cite{quercia2015smelly, quercia2015digital}, to exploiting user mobility dynamics and street network patterns~\cite{zhong2015revealing, zhong2014detecting, louf14, louail2014mobile, ren2014predicting, toole2015path} in order to build new models and tools describing cities, the prospects for the smart cities of the future look promising.



\section{Accessing the Data}
The Foursquare dataset used in the present work is proprietary and cannot be directly shared by the authors. In case of interest, similar datasets could be accessed through the OpenStreetMap project https://www.openstreetmap.org/ where crowd-sourced data about points-of-interest and their addition time becomes available. 

\section{Acknowledgements}
We acknowledge the support of the EPSRC Project GALE (EP/K019392).

\small
\bibliographystyle{plain}
\bibliography{biblio}

\clearpage
\normalsize
\appendices
\section{Identification of new places} \label{app:new_places}

This paper analyses the impact of new places in cities. However a precursor to any such analysis is the ability to identify which Foursquare venues are new places. As Foursquare was launched in 2009 with an empty database of venues, for any given venue the exact meaning of the \emph{date added} field is unknown. It may represent the approximate date the place opened, or it may be that the place pre-dates Foursquare and it represents the first visit of Foursquare user.

\begin{figure}[ht]
    \centering
    \includegraphics[width=0.8\columnwidth]{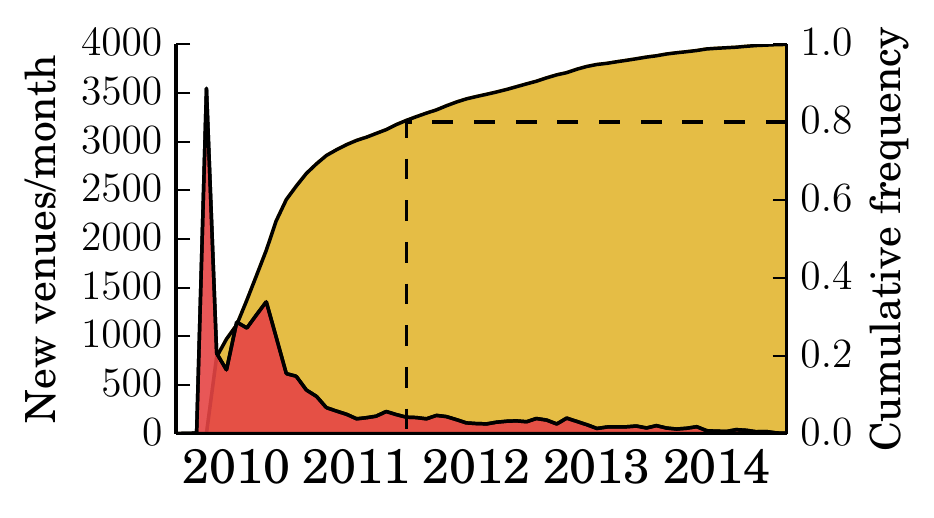}
    \caption{Venues by added per month. The dotted line represents our cut-off point between existing and new venues.}
    \label{fig:London_dates_added}
\end{figure}

\begin{figure}[ht]
    \centering
   	\includegraphics[width=0.8\columnwidth]{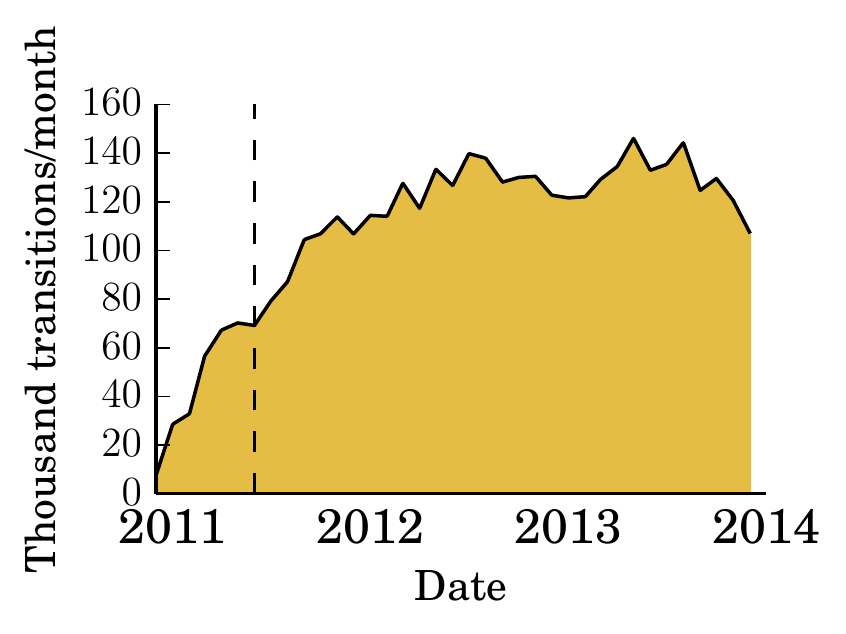}
    \caption{Foursquare's changing popularity as measured by transitions per month. The dotted line represents our cut-off point between existing and new venues.}
    \label{fig:London_transitions_per_month}
\end{figure}

To help overcome this we assume that, at some point in time, Foursquare adoption was sufficiently large that only a small fraction of venues remain unvisited by Foursquare users.  Figure~\ref{fig:London_dates_added}
shows the rate of new places being added to the London dataset over time and Figure~\ref{fig:London_transitions_per_month} shows how the number of transitions in London varies over time.

As expected, a large proportion of all venues were added to the dataset at the beginning of the period studied. Under our hypothesis these are existing places being added. Subsequently the number of new venues added decays sharply. After studying data from the other cities in the dataset, we conclude that the stated hypothesis seems reasonable. We found empirically that across nearly all cities the last 20\% of venues added to the database are contained in the comparatively flat tail of the distribution. Therefore we take the date where the number of new venues reaches 80\% of the city's total to be the cut-off point between new and existing venues. This point in time, assuming it exists, will be city-dependant.

Unfortunately the rate of new venues added does still gradually decrease after these cut-off points. Lacking any credible evidence that the growth rate of new places in cities is decreasing worldwide, we are forced to accept that we are still capturing some residual existing venues amongst the genuinely new venues. Unfortunately we have to accept this as a limitation of the dataset. We argue that the majority of venues we capture are indeed new venues as:
\begin{itemize}
\item Figure~\ref{fig:London_transitions_per_month} shows that Foursquare usage remains roughly constant in London from 2012 onward. Therefore the venues added after the cut-off date are not simply unpopular existing venues found by a rapidly growing user-base.

\item As mentioned in Section~\ref{sec:dataset}, in order to have been included in the dataset a venue must have had at least 100 check-ins. The greater a venue's total check-ins, the greater the probability of it being a genuinely new venue, as the likelihood of it not having been previously visited by a Foursquare user is increasingly remote. For example, in order for an existing venue to be added in 2013 no Foursquare user could ever have checked-in beforehand, but over 100 users subsequently checked-in from 2013--2014. While not impossible, this does not seem likely to occur frequently.

\item Finally in order to provide a rough estimate of our accuracy, we selected 50 random new places from our list of new places in London. For each of these we attempted to manually ascertain whether they were genuinely new places.

\begin{table}[ht]
	\centering
	\begin{tabular}{ccc}
		\toprule
		Genuinely new 	& Unverifiable 	& Existing places \\
		\midrule
		32				& 14			& 4 \\
		\bottomrule
	\end{tabular}
	\caption{Manual verification of 50 "new places"}
	\label{tab:50_new_places}
\end{table}

As can be seen in Table~\ref{tab:50_new_places}, only 4 places (less than 10\% of the sample) turned out to be existing places and two of these were a restaurant and a hotel that had closed for several years to undergo lengthy refurbishment and therefore could be argued to be new in the context of our analysis looking at the effects of places opening up. 13 out of the 14 unverified places were chain stores (Starbucks, River Island etc.) for which it is very difficult to find the precise opening date of an individual store. Given the turn-over of such chain stores, it is easy to believe that the majority of them were indeed new.
\end{itemize}

\end{document}